\documentclass[%
 reprint,
superscriptaddress,
prl,
 amsmath,amssymb,
 aps,
]{revtex4-2}

\usepackage{graphicx}
\usepackage{dcolumn}
\usepackage{bm}
\usepackage{xcolor}
\usepackage{physics, amsmath, amssymb}
\usepackage{bbold} 
\usepackage{hyperref}

\newcommand{\rev}[1]{{#1}}
\definecolor{blue}{rgb}{0,0,0}

\begin{document}

\preprint{APS/123-QED}

\title{Prodiabatic Elimination:\\ Higher Order Elimination of Fast Variables with Quantum Noise}

\author{Jan Neuser}
    \affiliation{Department of Physics, University of Basel, Klingelbergstrasse 82, CH-4056 Basel, Switzerland}
    \affiliation{Atominstitut, TU Wien, 1020 Vienna, Austria}
 \author{Marcelo Janovitch}
   \affiliation{Department of Physics, University of Basel, Klingelbergstrasse 82, CH-4056 Basel, Switzerland}
 \author{Matteo Brunelli}
   \affiliation{JEIP, UAR 3573 CNRS, Coll\`ege de France, PSL Research University}
 \author{Patrick P. Potts}
   \affiliation{Department of Physics, University of Basel, Klingelbergstrasse 82, CH-4056 Basel, Switzerland}
\date{\today}

\begin{abstract}
\rev{
Adiabatic elimination plays a central role in cavity quantum electrodynamics by removing fast cavity degrees of freedom, but at the cost of losing their finite-time response in multi-time correlations. We introduce prodiabatic elimination, a systematic expansion of the quantum Langevin equations that retains vacuum-noise contributions alongside higher-order corrections to the eliminated cavity field. This provides controlled access to arbitrary time- and normally ordered multi-time correlation functions, including photodetection observables, beyond what can be recovered from effective atomic dynamics alone. In the driven-dissipative Jaynes–Cummings model, we identify a vacuum-mediated effective repulsion between detected photons that suppresses short-delay bunching. For a single-photon switch, we analytically resolve a finite heralded response time arising from the cavity field’s delayed adjustment to the atomic transfer.
}
\end{abstract}
\maketitle
\textit{Introduction—}Exact descriptions of open quantum dynamics are rarely available, making approximation methods essential for constructing effective models. Systems exhibiting well-separated timescales are particularly amenable to such approaches, as the separation can be exploited to derive simplified effective models~\cite{bender:book}, with applications in physics~\cite{Rice:1988,Brion:2007,annby-andersson:2022}, chemistry~\cite{Gunawardena:2014}, biology~\cite{Multitimescales_in_biology}, medicine~\cite{Jard_n_Kojakhmetov_2021}, and even finance~\cite{Finance_Timescales}. One of the most prominent approaches to tackle systems with a separation of timescales is provided by adiabatic elimination~\cite{gardiner:1984,Rice:1988,PhysRevA.29.1438,PhysRevA.109.032603,Brion:2007,Finkelstein-Shapiro2019Adiabatic, PhysRevA.111.052206,10886784}, in which fast degrees of freedom are systematically removed, yielding an effective description for the remaining slow variables.

{\color{blue}
In cavity~\cite{RevModPhys.87.1379} and circuit quantum electrodynamics~\cite{CircuitQED}, optomechanics~\cite{CavityOptomechanics,ADBElim_quadratic_coupling}, and cavity materials engineering~\cite{Lu:25,Huebener2021,Mivehvar2021}, a rapidly decaying cavity mode is often coupled to a slower quantum system, which we generically call the atom, see Fig.~\ref{fig: sketch setup}. Adiabatic elimination then yields an effective description of the atom alone~\cite{Rice:1988}: at leading order in the separation of timescales, the cavity field is replaced by its instantaneous response to the atom and the drive. This is accurate for the slow atomic dynamics, but not for photon correlations at delays comparable to the cavity lifetime. When a photon is detected, the state of the atom is updated, and the cavity field needs about a cavity lifetime to adjust to it, which an instantaneous response cannot describe.

Beyond leading order, several approaches derive effective master equations for the atom~\cite{10886784,Azouit2017Towards,Marcuzzi:2014,Paulisch2014,CompletePostivityViolationADB, Marcelo_OneDim_Atom, giaccari2025}, with cavity vacuum noise entering only through the induced dissipation. Although this can capture corrections to the atomic dynamics, it does not recover the finite-time cavity response in photon correlations such as the second-order correlation function $g^{(2)}(\tau)$. The issue is that, despite vanishing from normally ordered expectation values, vacuum noise can contribute to multi-time correlations through its prior interaction with the atom. These contributions retain a memory of the cavity's finite response time. Accounting systematically for noise beyond leading order is notoriously subtle, even in classical stochastic systems~\cite{Sancho1982,vanKampen1985,GardinerHandbook2009WhiteNoise}.

\begin{figure}
    \centering
    \includegraphics[trim={29.9cm 11cm 7.5cm 13cm},clip,width=0.8\linewidth]
    {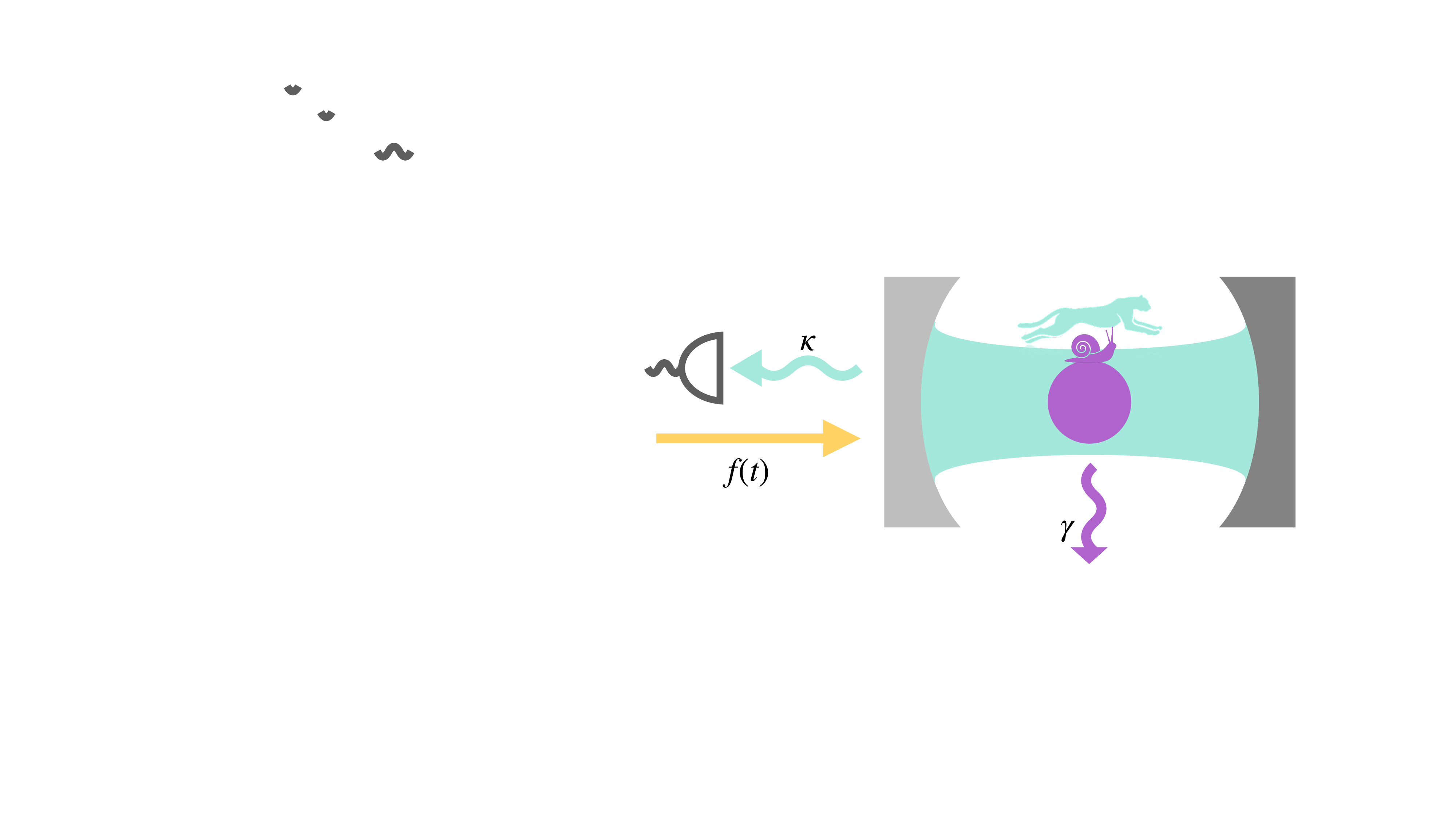}
    \caption{General setup. A quantum system (purple) is embedded in a driven cavity, {\color{blue} with measurement of dissipated light}. The quantum system is assumed to evolve slowly compared to the timescale governing the cavity, which allows the light to be eliminated. The cavity drive is denoted by $f(t)$, its linewidth by $\kappa$, and the quantum system dissipates at rate $\gamma$.}
    \label{fig: sketch setup}
\end{figure}

Here, we introduce \textit{prodiabatic elimination}, which eliminates the cavity from the quantum Langevin equations~\cite{InputOutputTheory} while retaining the vacuum-noise contributions needed for multi-time correlations. The method combines two ingredients: finite-response-time corrections to the cavity field, which also modify the effective atomic dynamics, and noise contractions that capture the cavity's conditional response following a detection. Multi-time photon correlations can thus be expressed in terms of atomic operators and explicit functions of the time delays, without retaining the cavity as a dynamical degree of freedom. For a zero-temperature environment, the approach retains much of the simplicity of leading-order adiabatic elimination.

We demonstrate the method in two examples. For a driven two-level atom in a cavity, prodiabatic elimination corrects the spurious nonzero initial slope of $g^{(2)}(\tau)$ predicted by adiabatic elimination and captures finite-linewidth effects in intensity correlations. For a single-photon switch based on a three-level atom, a detection in one cavity mode signals that the switch has closed, yet the field in the other mode takes at least a cavity lifetime to reach its new value. Even with corrected atomic dynamics, adiabatic elimination predicts an instantaneous switch; prodiabatic elimination instead captures the gradual recovery of the reflectance using atomic variables alone.
}

\textit{The prodiabatic elimination—}We model the dynamics of the cavity field with the quantum Langevin equation \cite{InputOutputTheory}\,($\hbar=1$)
\begin{equation}
    \dot{\hat{a}} = -\left(\frac{\kappa}{2} + i \Delta \right) \hat a - i g \hat b+ if(t)  - \sqrt{\kappa} \hat a_{\rm in} \, ,
    \label{eq: general dot a}
\end{equation}
where $\kappa$ is the cavity dissipation rate, $\Delta$ is the cavity frequency (or detuning in a rotating frame), $g$ denotes the coupling to the intracavity system which we generally refer to as atom, $f(t)$ is a (possibly time-dependent) coherent drive amplitude, and $\hat a_{\rm in}$ is an input operator of zero mean, describing the vacuum noise entering the cavity. The operator $\hat b$ acts on the atom and is assumed to be governed by the equation of motion

\begin{equation}
    \dot{\hat{b}} = -i\Omega\hat v -\hat{r}\left( \frac{\gamma}{2} \hat b + ig \hat a + \sqrt{\gamma} \hat b_{\rm in} \,\right) ,
      \label{eq: general dot b}
\end{equation}
where  $-i\Omega\hat v$ describes the evolution of $\hat b$ in the absence of the cavity and dissipation, $\gamma$ denotes the atomic dissipation rate, $\hat b_{\rm in}$ is a zero mean input operator, describing vacuum noise, and $\hat{r}=[\hat{b}, \hat{b}^\dagger]$. The terms proportional to $g$ in Eqs.~(\ref{eq: general dot a}, \ref{eq: general dot b}) result from a Jaynes--Cummings-like coupling $g(\hat a\hat b^\dagger+ \hat b \hat a^\dagger)$. A Markovian master equation that is equivalent to these Langevin equations is provided in the Supplemental Material~\cite{supmat}.

We introduce a small parameter $\epsilon$ which ensures a separation of timescales between the cavity, evolving fast, and the atom, evolving slowly;
\begin{align}
    \mathcal{O}\left(\frac{\gamma}{\kappa}\right) = \mathcal{O}\left(\frac{\Omega}{\kappa}\right) =\epsilon^2 \, , && \mathcal{O}\left(\frac{g}{\kappa} \right)  =  \mathcal{O}\left(\frac{f}{\kappa} \right)  =   \epsilon  \, .
    \label{eq: Epsilon orders of terms}
\end{align}
The conditions, apart from $\Omega/\kappa$, are known as \textit{the bad-cavity limit} \cite{Rice:1988}, implying the Purcell factor $F_{\rm p} = \frac{4g^2}{\gamma \kappa}$ to be of order $1$. 

Our goal is to evaluate time- and normally ordered expectation values of the form
\begin{equation}
        \mathcal{G}(\{\tau_m,t_n\}) =  \expval{ \hat a^\dagger(\tau_1) \cdots \hat a^\dagger(\tau_{M}) \hat \Sigma(t) \hat a(t_{N}) \cdots \hat a(t_1)} \, ,
     \label{eq: sigma in time ordered expecvalue}
 \end{equation}
where $\hat \Sigma(t)$ is an unspecified atom operator and time ordering means $t_1 \leq \cdots \leq t_{N}$ as well as $\tau_1 \leq \cdots \leq \tau_{M}$. {\color{blue} For $\hat\Sigma=\mathbb 1$, these correlations determine normally ordered photodetection observables~\cite{KelleyKleiner1964,Glauber1963,MandelWolf1995}.}

{\color{blue}
We start from the formal solution of Eq.~\eqref{eq: general dot a} with initial time in the far past,
\begin{equation}
\hat a(t)=-ig(\chi_{\rm c}*\hat b)(t)+iF(t)+\hat A_{\rm in}(t),
\label{eq: formal solution to general poblem}
\end{equation}
where $\chi_{\rm c}(\tau)=e^{-\kappa\tau/(2t_{\rm c})}$ for $\tau\geq0$ is the cavity response kernel with
response factor $t_{\rm c}=(1+2i\Delta/\kappa)^{-1}$, and 
$(\chi_{\rm c}*\hat O)(t)=\int_0^\infty d\tau\,\chi_{\rm c}(\tau)\hat O(t-\tau)$.
The second and third terms are the filtered drive  $F=\chi_{\rm c}*f$ and the filtered input noise $\hat A_{\rm in}=-\sqrt{\kappa}\,\chi_{\rm c}*\hat a_{\rm in}$, respectively. For a constant drive, $F=2t_{\rm c}f/\kappa$. At leading order, replacing the slowly evolving atomic operator inside the convolution by its instantaneous value and discarding input noise gives the standard adiabatic solution
\begin{equation}
\hat a_{\rm adb}(t)=-\frac{2it_{\rm c}}{\kappa}[g\hat b(t)-f(t)].
\end{equation}
To capture finite-bandwidth corrections and reconstruct multi-time photon correlations at higher orders, however, we must retain both the time dependence inside the convolution and the input-noise contributions.

We therefore expand the slowly evolving atomic operator inside the convolution and get the prodiabatic expansion
\begin{equation}
\hat a(t)\simeq
\hat a_{\rm pdb}(t)+\hat A_{\rm in}(t)
-2\sqrt{\kappa}\,\hat B(t)
\left(K*\hat a_{\rm in}^\dagger\right)(t),
\label{eq: general a_pdb with noise}
\end{equation}
which is our first main result. Besides the filtered input noise $\hat A_{\rm in}$, the field contains two contributions with distinct physical roles. The first, $\hat a_{\rm pdb}$, incorporates finite-time corrections to the instantaneous cavity field. The second 
describes the vacuum-mediated contribution needed to retain the cavity's finite response in multi-time correlations.

The first contribution is an atomic operator that retains terms through order $\epsilon^3$:
\begin{equation}
\begin{aligned}
\hat a_{\rm pdb}(t)={}&
\left[1+4\frac{g^2t_{\rm c}^2}{\kappa^2}\hat r(t)\right]
\left[-\frac{2igt_{\rm c}}{\kappa}\hat b(t)+iF(t)\right]\\
&-2i\frac{\gamma gt_{\rm c}^2}{\kappa^2}\hat r(t)\hat b(t)
+4\frac{\Omega gt_{\rm c}^2}{\kappa^2}\hat v(t).
\end{aligned}
\label{eq: apdb without noise}
\end{equation}
It provides a corrected atomic representation of the cavity field, accounting for its finite response time and the associated corrections to the effective atomic dynamics.

The second contribution is governed by the atomic operator
\begin{equation}
\hat B= \frac{4t_{\rm c}^3 g^2}{\kappa^3} \left(
i\Omega[\hat v,\hat b]
+F(t)g[\hat b,\hat r]
-\frac{\gamma}{2}(1+F_{\rm p}t_{\rm c})[\hat b,\hat r]\hat b
\right),
\label{eq:Bop}
\end{equation}
which is of order $\epsilon^4$. It multiplies the filtered input creation operator through the modified response kernel
\begin{equation}
K(\tau)=
\frac{t_{\rm c}^*[\chi_{\rm c}^*(\tau)-\chi_{\rm c}(\tau)]}
{t_{\rm c}^*-t_{\rm c}},
\label{eq: response kernel}
\end{equation}
whose resonant limit is $K(\tau)=(\kappa\tau/2)e^{-\kappa\tau/2}$.
It determines the temporal profile of the vacuum-mediated contribution: the input creation operator $\hat a_{\rm in}^\dagger$ is filtered by $K$ and multiplied by $\hat B(t)$.

Although the input is in vacuum, this term cannot be discarded when evaluating multi-time correlations. The filtered input noise has a nontrivial commutator with the cavity field because the input has previously interacted with the atom through the cavity. We find that $[\hat A_{\rm in}(t),\hat a(t')]
=\chi_{\rm c}(t-t')\hat B(t')+\mathcal O(\epsilon^5)$ with $ t\geq t'$. The resulting vacuum contractions generate contributions to normally ordered, multi-time correlations that retain the cavity's finite response time. The derivation of Eqs.~\eqref{eq: general a_pdb with noise}--\eqref{eq:Bop} is given in the Supplemental Material~\cite{supmat}.

Substituting Eq.~\eqref{eq: general a_pdb with noise} into Eq.~\eqref{eq: sigma in time ordered expecvalue} and evaluating the vacuum contractions 
we obtain the  multi-time correlations
}
\begin{widetext}
\begin{equation}
\begin{aligned}
    &\mathcal{G}\left(\{\tau_m, t_n\}\right) = \bigg\langle \hat{a}_{\rm pdb}^\dagger(\tau_1)\cdots \hat{a}_{\rm pdb}^\dagger(\tau_M) \hat{\Sigma}(t) \hat{a}_{\rm pdb}(t_{N}) \cdots  \hat{a}_{\rm pdb}(t_1)\bigg\rangle\\
    &+ \sum_{i = 1 }^{N-1}  \sum_{j = i +1}^{N}  \rev{\chi_{\rm c}(t_j-t_i)}\bigg\langle\mathcal{T}_\mathcal{G} \hat{a}_{\rm pdb}^\dagger(\tau_1)\cdots \hat{a}_{\rm pdb}^\dagger(\tau_M)\hat{\Sigma}(t)\hat{B}(t_i)\prod_{n\neq i,j}^N \hat{a}_{\rm pdb}(t_n)\bigg\rangle\\
    &+ \sum_{i = 1 }^{M-1}  \sum_{j = i +1}^{M} \rev{\chi_{\rm c}^*(\tau_j-\tau_i)}\bigg\langle\mathcal{T}_\mathcal{G}\hat{B}^\dagger(\tau_i)\bigg[\prod_{k\neq i,j}^M \hat{a}_{\rm pdb}^\dagger(\tau_k)\bigg]\hat{\Sigma}(t)  \hat{a}_{\rm pdb}(t_{N}) \cdots  \hat{a}_{\rm pdb}(t_1)\bigg\rangle \, +\mathcal{O}(\epsilon^{N+M+3}),
\end{aligned}
\label{eq: General G with noise}
\end{equation}
\end{widetext}
{\color{blue}
which is the second main result of this work. Here $\mathcal T_{\mathcal G}$ ensures time ordering and restores the ordering in Eq.~\eqref{eq: sigma in time ordered expecvalue}. 
The first line contains correlations of the corrected atomic field, while the second and third lines describe the vacuum-mediated cavity response through the kernel $\chi_{\rm c}$. These terms retain the cavity's finite response time in multi-time correlations without retaining the cavity as a dynamical degree of freedom. Their explicit kernels decay on the scale of the cavity lifetime.

\begin{figure}[t]
 \centering
 \includegraphics[width=\columnwidth]{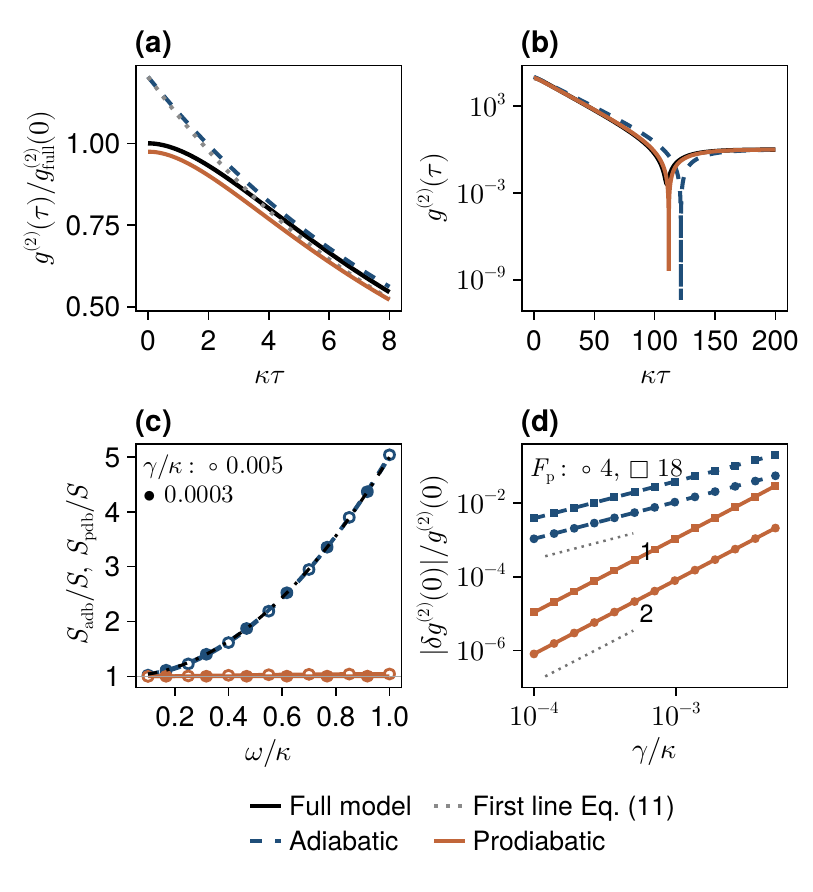}
 \caption{\rev{Conditional cavity response in the driven Jaynes--Cummings model. Black: numerical simulation of the full master equation; blue dashed: adiabatic; grey dotted: first line of Eq.~\eqref{eq: General G with noise} only; orange: prodiabatic. (a) Short-delay correlations, all divided by the same full-model $g^{(2)}(0)$. (b) Longer-delay minimum; black uses finite drive, colored curves the weak-drive formulas. In (a,b), $F_{\rm p}=18$, $\gamma/\kappa=0.005$, $f/\kappa=2.5\times10^{-4}$ and $\Delta=\Omega=0$. (c) Signed ratios of the adiabatic and prodiabatic spectra, $S_{\rm adb}$ and $S_{\rm pdb}$, to the exact $S$ at $F_{\rm p}=4$, for $\gamma/\kappa=0.005$ (open circles) and $0.0003$ (filled circles); the black dash-dotted line is $1+(2\omega/\kappa)^2$. (d) Relative zero-delay errors for $F_{\rm p}=4$ (circles) and $18$ (squares); grey guides indicate first and second order. Panels (c,d) use the weak-drive limit, checked against the full model.}}
 \label{fig: three figures}
\end{figure}


\textit{Photon correlation functions—}For correlation functions involving at most one cavity creation and one annihilation operator, only the first line of Eq.~\eqref{eq: General G with noise} contributes at the order considered. In particular, the reduced equations for atomic observables follow from the ordered atomic equations. The pair sums vanish whenever $N\leq1$ and $M\leq1$, but contribute to higher-order correlations, including equal-time moments, when $N\geq2$ or $M\geq2$; we refer to these contributions as \emph{pair terms}. They are therefore essential for observables derived from higher-order correlations, such as the intensity and squeezing spectra.

To illustrate their effect, consider 
\begin{equation}
 S(\omega)=2\int_0^\infty dt\,\cos(\omega t)[g^{(2)}(t)-1],
 \label{eq: excess spectrum}
\end{equation}
where $g^{(2)}(t)=\langle\hat a^\dagger(0)\hat a^\dagger(t)\hat a(t)\hat a(0)\rangle/\langle\hat a^\dagger(0)\hat a(0)\rangle^2$.  $S(\omega)$ describes the spectrum of \textit{intensity} fluctuations in excess of the shot-noise background. On cavity resonance, at frequencies above the slow atomic frequencies and decay rates, the pair terms yield~\cite{supmat}
\begin{equation}
 S(\omega)\simeq[1+(2\omega/\kappa)^2]^{-1}S_{\rm adb}(\omega),
 \label{eq: filter law}
\end{equation}
provided the leading spectral tail is nonzero. Therefore, even when the cavity is eliminated from the dynamics, its finite response leaves a frequency-dependent signature in the intensity spectrum: the high-frequency tail is filtered by the cavity response. This correction persists as the scale separation increases at fixed $\omega/\kappa=\mathcal O(1)$, but disappears in the slow-frequency limit. The approximation is not uniform near spectral cancellations or at arbitrarily high frequencies. 
This result is general for the resonant setup considered here; its consequences, together with other implications of Eq.~\eqref{eq: General G with noise}, are illustrated in the examples below.

The pair terms also show why the general result cannot be reproduced simply by replacing $\hat a\to\hat a_{\rm pdb}$ in a Lindblad master equation and applying the quantum regression theorem. Such a prescription captures the effective atomic contribution but misses the explicit cavity-response terms required for general multi-time correlations, even at the controlled order of the expansion.

\textit{Jaynes--Cummings model—}We first take $\hat b=\hat v=\hat\sigma=\ket0\!\bra1$ and $\hat r=-\hat\sigma_z$, where $\hat\sigma_z=\ket1\!\bra1-\ket0\!\bra0$. The constant-drive model and the reduced atomic equations are specified in the Supplemental Material and End Matter. On resonance, the atomic population decay rate is $\Gamma_{\rm adb}=\gamma(1+F_{\rm p})$ at leading order and $\Gamma_{\rm pdb}=\Gamma_{\rm adb}(1+\gamma F_{\rm p}/\kappa)$ after the prodiabatic correction; the coherence decays at half this rate.
For $\Delta=\Omega=0$ and $f^2/(\kappa\gamma)\ll1$, Eq.~\eqref{eq: General G with noise} gives 
\begin{align}
 g^{(2)}_{\rm pdb}(t)={}&\left(1-F_{\rm p}^2e^{-\Gamma_{\rm pdb}t/2}\right)^2\nonumber\\
 &+2e^{-\kappa t/2}\frac{\gamma}{\kappa}F_{\rm p}^2(1+F_{\rm p})
 [1-F_{\rm p}^2e^{-\Gamma_{\rm pdb}t/2}].
 \label{eq: g2 first order in f}
\end{align}
The second line is the pair-term correction. Adiabatic elimination omits it and replaces $\Gamma_{\rm pdb}$ by $\Gamma_{\rm adb}$. Corrections to the slow amplitude $F_{\rm p}^2$ enter beyond the retained order and are omitted consistently~\cite{supmat}.


Figures~\ref{fig: three figures}(a,b) resolve the cavity-time response.
The exact stationary correlation has zero initial slope; the pair term cancels the
leading slope of the first-line-only result. It also improves the relative zero-delay
error $|\delta g^{(2)}(0)|/g^{(2)}(0)$, where $\delta g^{(2)}(0)$ is the deviation of the adiabatic or prodiabatic result from the exact $g^{(2)}(0)$, 
changing its scaling from first to second order in $\Gamma_{\rm adb}/\kappa$, away from zeros of the
leading correlation [Fig.~\ref{fig: three figures}(d)]. 
The corrected atomic rate
independently improves the longer-delay minimum: the finite-drive full model gives
$\kappa t\simeq110.1$, compared with $121.7$ for adiabatic and $111.7$ for prodiabatic
elimination, Fig.~\ref{fig: three figures}(b). These correspond to errors of $10.5\%$ and $1.4\%$.

Both exact numerics and prodiabatic elimination indicate that at zero delay the photons are less bunched
than adiabatic elimination predicts. Prodiabatic elimination also identifies
the origin of this effect: after the first photon is detected, 
vacuum fluctuations filtered by the cavity response delay the second photon,
as if mediating a short-range repulsive force between the detected pair.

The spectral comparison in Fig.~\ref{fig: three figures}(c) tests Eq.~\eqref{eq: filter law}.
The fast contribution cancels part of the slow spectral tail. It need not be positive, so the comparison retains the sign of the excess spectrum. 
Further details
are provided in the Supplemental Material~\cite{supmat}.
}

{\color{blue}

\begin{figure}[t]
  \centering
  \includegraphics[width=\linewidth]{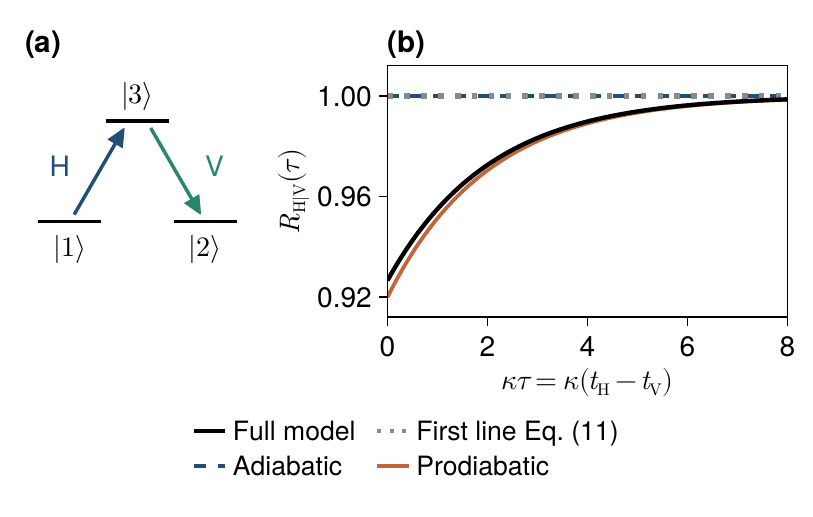}
  \caption{\rev{Photon switch. (a) H drives $\ket{1}\leftrightarrow\ket{3}$; a Raman photon in V transfers the atom to $\ket{2}$. (b) Conditional reflectance, Eq.~\eqref{eq: switch reflectance}, each curve against its own closed-switch response. Adiabatic elimination and the first line---Eq.~\eqref{eq: General G with noise} without $\hat B_{\rm HV}$---coincide at one. Parameters: $g/\kappa=0.1$, $\gamma/\kappa=5\times10^{-4}$ ($F_{\rm p}=80$), $\sigma\kappa=55$ and $\bar n=0.1$.}}
  \label{fig: lambda figures}
\end{figure}

\textit{Herald delay in a single-photon switch—}We consider a three-level atom in a $\Lambda$ configuration: An H-polarized cavity couples $\ket{1}\leftrightarrow\ket{3}$
and a V-polarized cavity couples $\ket{2}\leftrightarrow\ket{3}$, Fig.~\ref{fig: lambda figures}(a).
A weak resonant pulse drives H with the atom initially in $\ket{1}$. Emission into V (Raman) transfers the atom to $\ket{2}$; therein the H mode is effectively decoupled from the atom. Detecting the Raman photon thus \textit{heralds} the reflection of whatever H-polarized light remains—the multi-photon components of the incident pulse. This configuration can be understood as \textit{a single-photon switch}~\cite{Shomroni2014,Rosenblum2016,DuanKimble2004,Aoki2009,OShea2013,Tiecke2014}.

For an ideal switch, the H port does not reflect before a single photon is measured in the V port, $R_{\rm H}=0$ (open switch); a V detection heralds the transfer $\ket{1}\to\ket{2}$, after which the conditional H-port reflectance is $R_{\rm H|V}=1$ (closed switch). We emphasize that \textit{switching} is conditioned on heralding the Raman photon.
Adiabatic elimination predicts heralding to be an instantaneous process at the Raman detection time. We show that the conditional reflectance instead has a transient deficit, $R_{\rm H|V}(\tau)=1-\delta(\tau)$, where $\tau>0$ is the time after the Raman detection.
Prodiabatic elimination identifies the conditional H field left by the atom as the source of $\delta(\tau)$ and predicts its relaxation. This relaxation originates from the delay necessary for heralding to propagate, which we call  \textit{herald delay}. 

We take equal couplings $g$, linewidths $\kappa$ and decay rates $\gamma$ on
both transitions, with no repump. Here $\Gamma_{\rm adb}=\gamma(1+F_{\rm p})$ is the leading atomic coherence decay rate, and $\Gamma_{\rm pdb}=\Gamma_{\rm adb}(1+2\gamma F_{\rm p}/\kappa)$ its prodiabatic value. For two modes, the pair operator carries channel indices. Its off-diagonal element $\hat B_{\rm HV}$ reconstructs the H field conditioned on a Raman detection~\cite{supmat}.

The detected fields are $\hat c_{\rm H}=\sqrt\kappa\hat a_{\rm H}-if(t)/\sqrt\kappa$ and $\hat c_{\rm V}=\sqrt\kappa\hat a_{\rm V}$; the H output includes interference with the coherent input. After a Raman detection at time $t$, the atom remains in $\ket{2}$ while H evolves as a driven empty cavity from the conditional field left at detection. With $\langle\,\cdot\,\rangle_{\rm V}$ denoting the average conditioned on that detection, Eq.~\eqref{eq: General G with noise} gives, to the retained order~\cite{supmat},
\begin{equation}
 \langle\hat a_{\rm H}(t+\tau)\rangle_{\rm V}\simeq iF(t+\tau)-i\,\frac{\gamma F_{\rm p}}{2\kappa}\,\chi_{\rm c}(\tau)F(t)
 \label{eq: conditional field}
\end{equation}
for a delay $\tau\geq0$. The conditional deficit in the second term relaxes over $2/\kappa$.

The conditional reflectance is the reflected H flux after a Raman detection at time $t$, normalized to the closed-switch flux of the same pulse and averaged over $t$,
\begin{align}
 R_{\rm H|V}(\tau)&=\frac{\int\!dt\,\Phi_{\rm V}(t)\,\langle\hat c_{\rm H}^\dagger\hat c_{\rm H}(t+\tau)\rangle_{\rm V}}
 {\int\!dt\,\Phi_{\rm V}(t)\,\Phi_{\rm H}^{\rm cl}(t+\tau)}
 =1-\delta(\tau),
 \label{eq: switch reflectance}
\end{align}
with the Raman flux $\Phi_{\rm V}(t)=\langle\hat c_{\rm V}^\dagger\hat c_{\rm V}(t)\rangle$ and the closed-switch flux $\Phi_{\rm H}^{\rm cl}(t)=\kappa\left[F(t)-f(t)/\kappa\right]^{2}$.

The reflectance deficit $\delta(\tau)$, averaged over the Raman detection time as in a coincidence measurement, is given by prodiabatic elimination,
\begin{equation}
 \delta(\tau)\simeq\frac{2\gamma F_{\rm p}}\kappa e^{-\kappa\tau/2}\mathcal{Q}_{F}(\tau),
 \label{eq: reflectance deficit}
\end{equation}
where $\mathcal{Q}_{F}(\tau)$ is generally  non-exponential since it depends on convolutions involving the filtered drive $F(t)$ and $\langle \hat\sigma_{33}(t)\rangle$~\cite{supmat}; this means a herald delay beyond a cavity lifetime due to the atom lagging the drive, and Raman photons arriving while the drive decreases.
If $F(t)\simeq \text{const.}$ within the atomic response time $\Gamma_{\rm pdb}^{-1}$, we have $\mathcal{Q}_{F}\simeq 1$, and the herald delay is the cavity lifetime $2/\kappa$. }
{\color{blue} 


Adiabatic elimination gives $R_{\rm H|V}\equiv1$, since it assigns the empty-cavity field as soon as the atom switches.
So does the first line of Eq.~\eqref{eq: General G with noise}, even with corrected atomic dynamics and output operators. The pair term reconstructs the missing conditional field, recovering the transient without retaining the cavity Hilbert space. Importantly, this shows that the cavity implementation—highly desirable due to the Purcell effect—also imposes a limitation on how fast the switch can operate: the herald delay originating from vacuum-noise-mediated back-action.

For experimental scale, reported parameters give $2\gamma F_{\rm p}/\kappa\simeq0.4\%$ for a nanophotonic cavity~\cite{Tiecke2014} and $\sim35\%$ for fiber-coupled microspheres used to route and extract single photons~\cite{Shomroni2014,Rosenblum2016}, although translating these rate ratios into device reflectances requires accounting for the different port geometries~\cite{supmat}.

Figure~\ref{fig: lambda figures}(b) tests this for a Gaussian pulse~\cite{supmat} at $\sigma\kappa=55$, where $\sigma$ is the rms width, and $\bar n=\int dt\,f(t)^2/\kappa=0.1$ incident photons, with the analytic form of $\mathcal{Q}_F(\tau)$ for this pulse; solving the atomic dynamics numerically changes the curve by less than $10^{-3}$. The full model returns $R_{\rm H|V}(0)=0.927$ and recovers by $6.8\%$ between $\tau=0$ and $4/\kappa$, against $0.920$ and $7.5\%$ for prodiabatic elimination. At zero delay, prodiabatic elimination thus reduces the error of the atom-only descriptions from $7.9\%$ to $0.7\%$, and the residual scales as $(\gamma F_{\rm p}/\kappa)^2$~\cite{supmat}. The End Matter relates $R_{\rm H|V}$ to the coincidence density that measures it, and shows that the recovery survives detector bins and timing jitter below $0.5/\kappa$.
}

\textit{Conclusion and outlook—}Prodiabatic elimination provides analytical expressions capturing a wealth of phenomena beyond the standard adiabatic elimination. This includes computing correlation functions beyond leading order, capturing short-time behavior via vacuum noise contributions, as well as the combined effect of pulse modulation and cavity filtering. 

Future extensions of the method may include treating thermal noise, as well as a general procedure to obtain Lindblad master equations.
Another interesting direction is the application of prodiabatic elimination to dispersive light-matter interactions in cavity QED, suitable for implementing quantum non-demolition measurement and spin squeezing, where higher-order corrections to adiabatic elimination are relevant to capture limitations to spin squeezing generation~\cite{Barberena2024, Caprotti2024, giaccari2025}.
\begin{acknowledgments}
\textit{Acknowledgments—}This work was supported by the Swiss National Science Foundation (Eccellenza Professorial Fellowship PCEFP2\_194268).
J.N. acknowledges support from the European Research Council (ERC Project 'Cocoquest' 101043705).
M.B. acknowledges funding from the European Research Council (ERC) under the European Union’s Horizon 2020 research and innovation program (Grant agreement No. 101002955 – CONQUER).
\end{acknowledgments}

\bibliography{References}

@misc{supmat,
  note = "In the {S}upplemental {M}aterial we provide details on the derivation of the prodiabatic elimination and its application to the considered examples."
}

@Article{Rice:1988,
  author   = {Rice, P.R. and Carmichael, H.J.},
  journal  = {IEEE J. Quantum Electron.},
  title    = {Single-Atom Cavity-Enhanced Absorption. I. Photon Statistics In The Bad-Cavity Limit},
  year     = {1988},
  number   = {7},
  pages    = {1351--1366},
  volume   = {24},
  doi      = {10.1109/3.974},
  fjournal = {IEEE Journal of Quantum Electronics},
}

@article{Sancho1982,
	author = {Sancho, J. M. and Miguel, M. San and D{\"u}rr, D.},
	date = {1982/06/01},
	doi = {10.1007/BF01012607},
	id = {Sancho1982},
	isbn = {1572-9613},
	journal = {J. Stat. Phys.},
	number = {2},
	pages = {291--305},
	title = {Adiabatic elimination for systems of {B}rownian particles with nonconstant damping coefficients},
	url = {https://doi.org/10.1007/BF01012607},
	volume = {28},
	year = {1982}
}

@Article{Marcuzzi:2014,
  author    = {Marcuzzi, M and Schick, J and Olmos, B and Lesanovsky, I},
  title     = {Effective dynamics of strongly dissipative {R}ydberg gases},
  journal   = {J. Phys. A: Math. Theor.},
  year      = {2014},
  volume    = {47},
  number    = {48},
  pages     = {482001},
  month     = {nov},
  doi       = {10.1088/1751-8113/47/48/482001},
  publisher = {IOP Publishing},
  url       = {https://doi.org/10.1088/1751-8113/47/48/482001},
}

@article{vanKampen1985,
	author = {N.G. {Van Kampen}},
	doi = {https://doi.org/10.1016/0370-1573(85)90002-X},
	issn = {0370-1573},
	journal = {Phys. Rep.},
	number = {2},
	pages = {69-160},
	title = {Elimination of fast variables},
	url = {https://www.sciencedirect.com/science/article/pii/037015738590002X},
	volume = {124},
	year = {1985}
}

@article{PhysRevA.109.032603,
  title = {Adiabatic elimination for composite open quantum systems: Reduced-model formulation and numerical simulations},
  author = {Le R\'egent, Francois-Marie and Rouchon, Pierre},
  journal = {Phys. Rev. A},
  volume = {109},
  issue = {3},
  pages = {032603},
  numpages = {20},
  year = {2024},
  month = {Mar},
  publisher = {American Physical Society},
  doi = {10.1103/PhysRevA.109.032603},
  url = {https://link.aps.org/doi/10.1103/PhysRevA.109.032603}
}

@article{Paulisch2014,
	author = {Paulisch, Vanessa and Rui, Han and Ng, Hui Khoon and Englert, Berthold-Georg},
	date = {2014/01/28},
	doi = {10.1140/epjp/i2014-14012-8},
	id = {Paulisch2014},
	isbn = {2190-5444},
	journal = {Eur. Phys. J. Plus},
	number = {1},
	pages = {12},
	title = {Beyond adiabatic elimination: A hierarchy of approximations for multi-photon processes},
	url = {https://doi.org/10.1140/epjp/i2014-14012-8},
	volume = {129},
	year = {2014}
}

@Book{bender:book,
  title     = {Advanced Mathematical Methods for Scientists and Engineers I - Asymptotic Methods and Perturbation Theory},
  publisher = {Springer New York, NY},
  year      = {1999},
doi = {10.1007/978-1-4757-3069-2},
  author    = {C. M. Bender and S. A. Orszag},
}

@Article{Gunawardena:2014,
  author        = {Gunawardena, Jeremy},
  title         = {Time-scale separation--{M}ichaelis and {M}enten's old idea, still bearing fruit},
  journal       = {FEBS J.},
  year          = {2014},
  volume        = {281},
  pages         = {473-88},
  month         = {Jan},
  doi           = {10.1111/febs.12532},
}

@article{annby-andersson:2022,
  title = {Quantum Fokker-Planck Master Equation for Continuous Feedback Control},
  author = {Annby-Andersson, Bj\"orn and Bakhshinezhad, Faraj and Bhattacharyya, Debankur and De Sousa, Guilherme and Jarzynski, Christopher and Samuelsson, Peter and Potts, Patrick P.},
  journal = {Phys. Rev. Lett.},
  volume = {129},
  issue = {5},
  pages = {050401},
  numpages = {8},
  year = {2022},
  month = {Jul},
  publisher = {American Physical Society},
  doi = {10.1103/PhysRevLett.129.050401},
  url = {https://link.aps.org/doi/10.1103/PhysRevLett.129.050401}
}

@incollection{GardinerHandbook2009WhiteNoise,
  author    = {Crispin Gardiner},
  title     = {The White Noise Limit},
  booktitle = {Stochastic Methods: A Handbook for the Natural and Social Sciences},
  publisher = {Springer Berlin Heidelberg},
  year      = {2009},
  edition   = {4},
  chapter   = {8},
  pages     = {185--214},
  address   = {Berlin, Heidelberg},
  isbn      = {978-3-540-70712-7},
}

@Article{Huebener2021,
  author        = {Hübener, Hannes and De Giovannini, Umberto and Schäfer, Christian and Andberger, Johan and Ruggenthaler, Michael and Faist, Jerome and Rubio, Angel},
  title         = {Engineering quantum materials with chiral optical cavities},
  journal       = {Nat. Mater.},
  year          = {2021},
  volume        = {20},
  number        = {4},
  pages         = {438--442},
  month         = apr,
  issn          = {1476-4660},
  refid         = {Hübener2021},
  url           = {https://doi.org/10.1038/s41563-020-00801-7},
}

@article{Lu:25,
author = {I-Te Lu and Dongbin Shin and Mark Kamper Svendsen and Simone Latini and Hannes H\"{u}bener and Michael Ruggenthaler and Angel Rubio},
journal = {Adv. Opt. Photon.},
number = {2},
pages = {441--525},
publisher = {Optica Publishing Group},
title = {Cavity engineering of solid-state materials without external driving},
volume = {17},
month = {Jun},
year = {2025},
url = {https://opg.optica.org/aop/abstract.cfm?URI=aop-17-2-441},
doi = {10.1364/AOP.544138},
}

@Article{Mivehvar2021,
  author    = {Farokh Mivehvar and Francesco Piazza and Tobias Donner and Helmut Ritsch and},
  title     = {Cavity QED with quantum gases: new paradigms in many-body physics},
  journal   = {Adv. Phys.},
  year      = {2021},
  volume    = {70},
  number    = {1},
  pages     = {1--153},
  doi       = {10.1080/00018732.2021.1969727},
  publisher = {Taylor \& Francis},
  url       = { 
    
        https://doi.org/10.1080/00018732.2021.1969727
    
    

},
}

@article{RevModPhys.87.1379,
  title = {Cavity-based quantum networks with single atoms and optical photons},
  author = {Reiserer, Andreas and Rempe, Gerhard},
  journal = {Rev. Mod. Phys.},
  volume = {87},
  issue = {4},
  pages = {1379--1418},
  numpages = {40},
  year = {2015},
  month = {Dec},
  publisher = {American Physical Society},
  doi = {10.1103/RevModPhys.87.1379},
  url = {https://link.aps.org/doi/10.1103/RevModPhys.87.1379}
}

@article{gardiner:1984,
  title = {Adiabatic elimination in stochastic systems. I. Formulation of methods and application to few-variable systems},
  author = {Gardiner, C. W.},
  journal = {Phys. Rev. A},
  volume = {29},
  issue = {5},
  pages = {2814--2822},
  numpages = {0},
  year = {1984},
  month = {May},
  publisher = {American Physical Society},
  doi = {10.1103/PhysRevA.29.2814},
  url = {https://link.aps.org/doi/10.1103/PhysRevA.29.2814}
}

@Article{InputOutputTheory,
  author    = {Gardiner, C. W. and Collett, M. J.},
  journal   = {Phys. Rev. A},
  title     = {Input And Output In Damped Quantum Systems: Quantum Stochastic Differential Equations And The Master Equation},
  year      = {1985},
  month     = jun,
  pages     = {3761--3774},
  volume    = {31},
  doi       = {10.1103/PhysRevA.31.3761},
  fjournal  = {Physical Review A: Atomic, Molecular, and Optical Physics},
  issue     = {6},
  numpages  = {0},
  publisher = {American Physical Society},
}

@Book{QuantumMasterEQ_Breuer,
  author    = {Breuer, Heinz-Peter and Petruccione, Francesco},
  publisher = {Oxford University Press},
  title     = {The Theory Of Open Quantum Systems},
  year      = {2007},
  isbn      = {9780199213900},
  month     = jan,
  doi       = {10.1093/acprof:oso/9780199213900.001.0001},
}

@article{Quantumregressiontheoremformulti-timecorrelators,
  title = {Quantum regression theorem for multi-time correlators: A detailed analysis in the {H}eisenberg picture},
  author = {Khan, Sakil and Agarwalla, Bijay Kumar and Jain, Sachin},
  journal = {Phys. Rev. A},
  volume = {106},
  issue = {2},
  pages = {022214},
  numpages = {16},
  year = {2022},
  month = {Aug},
  publisher = {American Physical Society},
  doi = {10.1103/PhysRevA.106.022214},
  url = {https://link.aps.org/doi/10.1103/PhysRevA.106.022214}
}

@Article{Marcelo_OneDim_Atom,
  author    = {Tomm, Natasha and Antoniadis, Nadia O. and Janovitch, Marcelo and Brunelli, Matteo and Schott, Rüdiger and Valentin, Sascha R. and Wieck, Andreas D. and Ludwig, Arne and Potts, Patrick P. and Javadi, Alisa and Warburton, Richard J.},
  journal   = {Phys. Rev. Lett.},
  title     = {Realization Of A Coherent And Efficient One-Dimensional Atom},
  year      = {2024},
  issn      = {1079-7114},
  month     = aug,
  number    = {8},
  pages     = {083602},
  volume    = {133},
  doi       = {10.1103/physrevlett.133.083602},
  fjournal  = {Physical Review Letters},
  publisher = {American Physical Society (APS)},
}

@Article{ADBElim_quadratic_coupling,
  author   = {Jiang, Cheng and Cui, Yuanshun and Chen, Guibin},
  journal  = {Sci. Rep.},
  title    = {Dynamics Of An Optomechanical System With Quadratic Coupling: Effect Of First Order Correction To Adiabatic Elimination},
  year     = {2016},
  number   = {1},
  pages    = {35583},
  volume   = {6},
  doi      = {10.1038/srep35583},
  fjournal = {Scientific Reports},
}

@Article{Brion:2007,
  author   = {Brion, E and Pedersen, L H and Mølmer, K},
  title    = {Adiabatic elimination in a lambda system},
  journal  = {J. Phys. A: Math. Theor.},
  year     = {2007},
  volume   = {40},
  number   = {5},
  pages    = {1033},
  month    = {jan},
  doi      = {10.1088/1751-8113/40/5/011},
  url      = {https://doi.org/10.1088/1751-8113/40/5/011},
}

@Article{PhysRevA.29.1438,
  author   = {L. Lugiato and P. Mandel and L. Narducci},
  journal  = {Phys. Rev. A},
  title    = {Adiabatic Elimination In Nonlinear Dynamical Systems},
  year     = {1984},
  pages    = {1438--1452},
  volume   = {29},
  doi      = {10.1103/PHYSREVA.29.1438},
  fjournal = {Physical Review A: Atomic, Molecular, and Optical Physics},
}

@Article{CircuitQED,
  author    = {Blais, Alexandre and Grimsmo, Arne L. and Girvin, S. M. and Wallraff, Andreas},
  journal   = {Rev. Mod. Phys.},
  title     = {Circuit Quantum Electrodynamics},
  year      = {2021},
  month     = may,
  pages     = {025005},
  volume    = {93},
  doi       = {10.1103/RevModPhys.93.025005},
  fjournal  = {Reviews of Modern Physics},
  issue     = {2},
  numpages  = {72},
  publisher = {American Physical Society},
}

@Article{CavityOptomechanics,
  author    = {Aspelmeyer, Markus and Kippenberg, Tobias J. and Marquardt, Florian},
  journal   = {Rev. Mod. Phys.},
  title     = {Cavity Optomechanics},
  year      = {2014},
  month     = dec,
  pages     = {1391--1452},
  volume    = {86},
  doi       = {10.1103/RevModPhys.86.1391},
  fjournal  = {Reviews of Modern Physics},
  issue     = {4},
  numpages  = {62},
  publisher = {American Physical Society},
}

@article{Finkelstein-Shapiro2019Adiabatic,
  title = {Adiabatic elimination and subspace evolution of open quantum systems},
  author = {Finkelstein-Shapiro, Daniel and Viennot, David and Saideh, Ibrahim and Hansen, Thorsten and Pullerits, T\~onu and Keller, Arne},
  journal = {Phys. Rev. A},
  volume = {101},
  issue = {4},
  pages = {042102},
  numpages = {14},
  year = {2020},
  month = {Apr},
  publisher = {American Physical Society},
  doi = {10.1103/PhysRevA.101.042102},
  url = {https://link.aps.org/doi/10.1103/PhysRevA.101.042102}
}

@article{Multitimescales_in_biology,
author = {Bertram, Richard and Rubin, Jonathan},
year = {2016},
month = {07},
pages = {},
title = {Multi-timescale Systems and Fast-Slow Analysis},
volume = {287},
journal = {Math. Biosci.},
doi = {10.1016/j.mbs.2016.07.003}
}

@article{Jard_n_Kojakhmetov_2021,
   title={A geometric analysis of the {SIR}, {SIRS} and {SIRWS} epidemiological models},
   volume={58},
   ISSN={1468-1218},
   DOI={10.1016/j.nonrwa.2020.103220},
   journal={Nonlinear Anal. Real World Appl.},
   publisher={Elsevier BV},
   author={Jardón-Kojakhmetov, Hildeberto and Kuehn, Christian and Pugliese, Andrea and Sensi, Mattia},
   year={2021},
   month=apr, pages={103220} }

@article{Finance_Timescales,
	author = {Ghashghaie, S. and Breymann, W. and Peinke, J. and Talkner, P. and Dodge, Y.},
	date = {1996/06/01},
	doi = {10.1038/381767a0},
	id = {Ghashghaie1996},
	isbn = {1476-4687},
	journal = {Nature},
	number = {6585},
	pages = {767--770},
	title = {Turbulent cascades in foreign exchange markets},
	url = {https://doi.org/10.1038/381767a0},
	volume = {381},
	year = {1996}
}

@article{Azouit2017Towards,
doi = {10.1088/2058-9565/aa7f3f},
url = {https://doi.org/10.1088/2058-9565/aa7f3f},
year = {2017},
month = {sep},
publisher = {IOP Publishing},
volume = {2},
number = {4},
pages = {044011},
author = {Azouit, R and Chittaro, F and Sarlette, A and Rouchon, P},
title = {Towards generic adiabatic elimination for bipartite open quantum systems},
journal = {Quantum Sci. Technol.}
}

@Article{CompletePostivityViolationADB,
  author    = {Tokieda, Masaaki and Elouard, Cyril and Sarlette, Alain and Rouchon, Pierre},
  journal   = {Phys. Rev. A},
  title     = {Complete positivity violation of the reduced dynamics in higher-order quantum adiabatic elimination},
  year      = {2024},
  month     = {Jun},
  pages     = {062206},
  volume    = {109},
  doi       = {10.1103/PhysRevA.109.062206},
  fjournal  = {Physical Review A: Atomic, Molecular, and Optical Physics},
  issue     = {6},
  numpages  = {25},
  publisher = {American Physical Society},
  url       = {https://link.aps.org/doi/10.1103/PhysRevA.109.062206},
}

@article{PhysRevA.111.052206,
  title = {Time-convolutionless master equation applied to adiabatic elimination},
  author = {Tokieda, Masaaki and Riva, Angela},
  journal = {Phys. Rev. A},
  volume = {111},
  issue = {5},
  pages = {052206},
  numpages = {20},
  year = {2025},
  month = {May},
  publisher = {American Physical Society},
  doi = {10.1103/PhysRevA.111.052206},
  url = {https://link.aps.org/doi/10.1103/PhysRevA.111.052206}
}

@INPROCEEDINGS{10886784,
  author={Riva, Angela and Sarlette, Alain and Rouchon, Pierre},
  booktitle={2024 IEEE 63rd Conference on Decision and Control (CDC)}, 
  title={Explicit formulas for adiabatic elimination with fast unitary dynamics}, 
  year={2024},
  volume={},
  number={},
  pages={755-760},
  doi={10.1109/CDC56724.2024.10886784}}

@article{giaccari2025,
 title={Higher-Order Adiabatic Elimination in Atom-Cavity Systems and Its Impact on Spin-Squeezing Generation},
 author={Giaccari, Stefano and Dellea, Giulia and Genoni, Marco G. and Bertaina, Gianluca},
 journal={Quantum Science and Technology},
 volume={11},
 pages={035019},
 year={2026},
 doi={10.1088/2058-9565/ae799f},
 eprint={2506.22383},
 archivePrefix={arXiv}
}

@article{Caprotti2024,
doi = {10.1088/2058-9565/ad4584},
url = {https://doi.org/10.1088/2058-9565/ad4584},
year = {2024},
month = {may},
publisher = {IOP Publishing},
volume = {9},
number = {3},
pages = {035032},
author = {Caprotti, A and Barbiero, M and Tarallo, M G and Genoni, M G and Bertaina, G},
title = {Analysis of spin-squeezing generation in cavity-coupled atomic ensembles with continuous measurements},
journal = {Quantum Science and Technology}
}

@article{Barberena2024,
  title = {Trade-offs between unitary and measurement induced spin squeezing in cavity QED},
  author = {Barberena, Diego and Chu, Anjun and Thompson, James K. and Rey, Ana Maria},
  journal = {Phys. Rev. Res.},
  volume = {6},
  issue = {3},
  pages = {L032037},
  numpages = {6},
  year = {2024},
  month = {Aug},
  publisher = {American Physical Society},
  doi = {10.1103/PhysRevResearch.6.L032037},
  url = {https://link.aps.org/doi/10.1103/PhysRevResearch.6.L032037}
}

@Article{KelleyKleiner1964,
  author  = {Kelley, P. L. and Kleiner, W. H.},
  journal = {Phys. Rev.},
  title   = {Theory of Electromagnetic Field Measurement and Photoelectron Counting},
  year    = {1964},
  issue   = {2A},
  pages   = {A316--A334},
  volume  = {136},
  doi     = {10.1103/PhysRev.136.A316},
}

@Article{Glauber1963,
  author  = {Glauber, Roy J.},
  journal = {Phys. Rev.},
  title   = {The Quantum Theory of Optical Coherence},
  year    = {1963},
  issue   = {6},
  pages   = {2529--2539},
  volume  = {130},
  doi     = {10.1103/PhysRev.130.2529},
}

@Book{MandelWolf1995,
  author    = {Mandel, L. and Wolf, E.},
  publisher = {Cambridge University Press},
  title     = {Optical Coherence and Quantum Optics},
  year      = {1995},
  address   = {Cambridge},
}

@article{Shomroni2014,
  author = {Shomroni, Itay and Rosenblum, Serge and Lovsky, Yulia and Bechler, Orel and Guendelman, Gabriel and Dayan, Barak},
  title = {All-Optical Routing of Single Photons by a One-Atom Switch Controlled by a Single Photon},
  journal = {Science},
  volume = {345},
  pages = {903--906},
  year = {2014},
  doi = {10.1126/science.1254699}
}

@article{Rosenblum2016,
  author = {Rosenblum, Serge and Bechler, Orel and Shomroni, Itay and Lovsky, Yulia and Guendelman, Gabriel and Dayan, Barak},
  title = {Extraction of a single photon from an optical pulse},
  journal = {Nature Photon.},
  volume = {10},
  pages = {19--22},
  year = {2016},
  doi = {10.1038/nphoton.2015.227}
}

@article{DuanKimble2004,
  author = {Duan, L.-M. and Kimble, H. J.},
  title = {Scalable Photonic Quantum Computation through Cavity-Assisted Interactions},
  journal = {Phys. Rev. Lett.},
  volume = {92},
  pages = {127902},
  year = {2004},
  doi = {10.1103/PhysRevLett.92.127902}
}

@article{Aoki2009,
  author = {Aoki, Takao and Parkins, A. S. and Alton, D. J. and Regal, C. A. and Dayan, Barak and Ostby, E. and Vahala, K. J. and Kimble, H. J.},
  title = {Efficient Routing of Single Photons by One Atom and a Microtoroidal Cavity},
  journal = {Phys. Rev. Lett.},
  volume = {102},
  pages = {083601},
  year = {2009},
  doi = {10.1103/PhysRevLett.102.083601}
}

@article{OShea2013,
  author = {O'Shea, Danny and Junge, Christian and Volz, J{\"u}rgen and Rauschenbeutel, Arno},
  title = {Fiber-Optical Switch Controlled by a Single Atom},
  journal = {Phys. Rev. Lett.},
  volume = {111},
  pages = {193601},
  year = {2013},
  doi = {10.1103/PhysRevLett.111.193601}
}

@article{Tiecke2014,
  author = {Tiecke, T. G. and Thompson, J. D. and de Leon, N. P. and Liu, L. R. and Vuleti{\'c}, V. and Lukin, M. D.},
  title = {Nanophotonic quantum phase switch with a single atom},
  journal = {Nature},
  volume = {508},
  pages = {241--244},
  year = {2014},
  doi = {10.1038/nature13188}
}

\clearpage
\appendix*
\setcounter{equation}{0}
\renewcommand{\theequation}{A\arabic{equation}}
\setcounter{figure}{0}
\renewcommand{\thefigure}{A\arabic{figure}}
\def\theHfigure{A\arabic{figure}}
\section{\rev{End Matter}}

{\color{blue}
\textit{Reduced Jaynes--Cummings dynamics—}The atomic equations contain $\langle\hat\sigma_z\hat a\rangle$ and $\langle\hat\sigma^\dagger\hat a\rangle$, so only the first line of Eq.~\eqref{eq: General G with noise} enters. With $t_{\rm q}=(1+2i\Omega/\gamma)^{-1}$, the reconstructed field is
\begin{equation}
 \hat a_{\rm pdb}=-\frac{2igt_{\rm c}}{\kappa}\left[\left(1+\frac{\gamma t_{\rm c}}{\kappa t_{\rm q}}+\frac{4g^2t_{\rm c}^2}{\kappa^2}\right)\hat\sigma
 +\frac{4gft_{\rm c}^2}{\kappa^2}\hat\sigma_z-\frac f g\right].
 \label{eq: a_p^1 in linearized form}
\end{equation}
The corresponding equations are
\begin{align}
 \langle\dot{\hat\sigma}\rangle&=-\frac{\Gamma_{\rm pdb}}2\langle\hat\sigma\rangle
 -\frac{2gft_{\rm c}}\kappa\left(\langle\hat\sigma_z\rangle-\frac{\gamma t_{\rm c}^2}\kappa F_{\rm p}\right),\label{eq:avgjc1}\\
 \langle\dot{\hat\sigma}_z\rangle&=-\Re\Gamma_{\rm pdb}(\langle\hat\sigma_z\rangle+1)\nonumber\\
 &\quad+\frac{8gf}\kappa\Re\left[t_{\rm c}\left(1+\frac{\gamma t_{\rm c}^2}\kappa F_{\rm p}\right)\langle\hat\sigma^\dagger\rangle\right],\label{eq:avgjc2}
\end{align}
where $\Gamma_{\rm pdb}=(\gamma/t_{\rm q})(1+t_{\rm c}t_{\rm q}F_{\rm p})(1+\gamma t_{\rm c}^2F_{\rm p}/\kappa)$. At resonance this is the real population rate used in the Letter; off resonance it also includes a frequency shift. The Supplemental Material gives a generator reproducing these equations through the retained order. Regression~\cite{Quantumregressiontheoremformulti-timecorrelators,QuantumMasterEQ_Breuer} with that generator and $\hat a_{\rm pdb}$ supplies the first line of Eq.~\eqref{eq: General G with noise}; its explicit pair terms must still be added.
}

\begin{figure}[t]
  \centering
  \includegraphics[width=\linewidth]{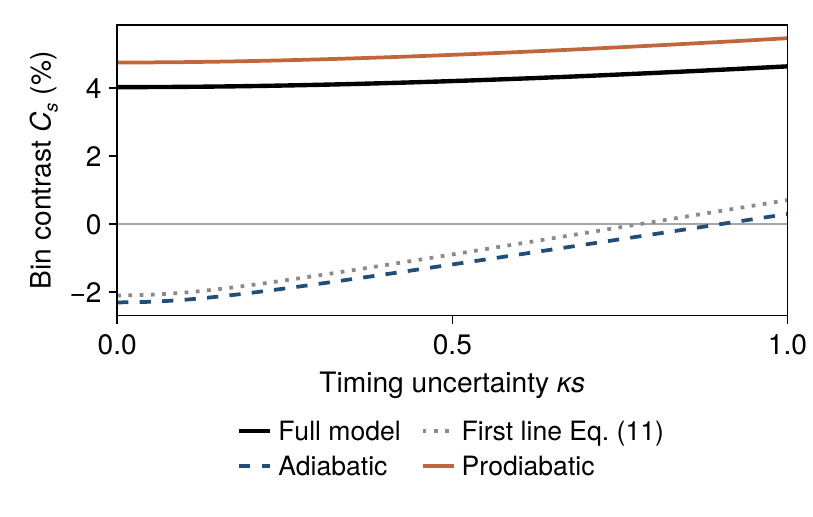}
  \caption{\rev{Contrast between the coincidence bins at zero delay and at $4/\kappa$, Eq.~\eqref{eq: bin contrast}, against the timing uncertainty of the delay. Line styles and parameters as in Fig.~\ref{fig: lambda figures}.}}
  \label{fig: switch detector}
\end{figure}

{\color{blue}
\textit{Coincidence density and herald yield—}The reflectance is measured through the coincidence density of a Raman photon and a reflected one at a delay $\tau\geq0$. By the definition of the conditional average, this density is the Raman rate times the conditional H intensity, integrated over the Raman detection time $t$,
\begin{align}
 P_{\rm VH}(\tau)&=\int\!dt\,\langle\hat c_{\rm V}^\dagger(t)\hat c_{\rm H}^\dagger(t+\tau)\hat c_{\rm H}(t+\tau)\hat c_{\rm V}(t)\rangle\nonumber\\
 &=\int\!dt\,\langle\hat c_{\rm V}^\dagger\hat c_{\rm V}(t)\rangle\,\langle\hat c_{\rm H}^\dagger\hat c_{\rm H}(t+\tau)\rangle_{\rm V}.
 \label{eq: switch profile}
\end{align}
We define the closed-switch coincidence density $P_{\rm VH}^{\rm cl}$ as the same integral with the conditional intensity replaced by the closed-switch intensity of Eq.~\eqref{eq: switch reflectance}: it is the coincidence density expected if the switch closed at the Raman detection. The reflectance is the ratio of the two,
\begin{align}
 P_{\rm VH}^{\rm cl}(\tau)&=\int\!dt\,\langle\hat c_{\rm V}^\dagger\hat c_{\rm V}(t)\rangle\,\kappa\Big[F(t+\tau)-\frac{f(t+\tau)}{\kappa}\Big]^{2},\nonumber\\
 R_{\rm H|V}(\tau)&=\frac{P_{\rm VH}(\tau)}{P_{\rm VH}^{\rm cl}(\tau)}.
 \label{eq: switch reference}
\end{align}
This normalizes the coincidences like a cross-correlation $g^{(2)}$. The ratio is Eq.~\eqref{eq: switch reflectance} averaged over the detection time, as Fig.~\ref{fig: lambda figures}(b) shows it, and equals Eq.~\eqref{eq: switch reflectance} for a slow drive.

The reference takes a simple approximate form for a Gaussian drive $f(t)=f_0\exp[-(t-t_0)^2/(2\sigma^2)]$ of amplitude $f_0$ centered at $t_0$ and carrying $\bar n$ incident photons, and for $\sigma\kappa\gg1$ and $\sigma\Gamma_{\rm pdb}\gg1$~\cite{supmat},
\begin{equation}
 P_{\rm VH}^{\rm cl}(\tau)\simeq\frac{N_{\rm V}\bar n}{\sqrt{2\pi}\,\sigma}\,
 \frac{e^{-\tau^2/(2\sigma^2)}}{\left[1+\tau/(2\sigma^2\Gamma_{\rm pdb})\right]^{2}},
 \label{eq: switch closed form}
\end{equation}
where $N_{\rm V}=\int dt\,\langle\hat c_{\rm V}^\dagger\hat c_{\rm V}\rangle\simeq\bar n\,(\gamma F_{\rm p}/\Gamma_{\rm adb})^2$ is the mean Raman count and $\Gamma_{\rm pdb}=\Gamma_{\rm adb}(1+2\gamma F_{\rm p}/\kappa)$ is the atomic coherence decay rate that prodiabatic elimination returns for the switch. The reference is the herald yield times the closed-switch flux that follows a herald. It varies only over the pulse and atomic times, so structure on the scale $1/\kappa$ belongs to $R_{\rm H|V}$.

The corrected atomic dynamics alone fix the herald yield and $P_{\rm VH}^{\rm cl}$; the pair term enters only through $R_{\rm H|V}$. The raw coincidences, on which the contrast below relies, rise with delay only for $\sigma\Gamma_{\rm adb}\gtrsim1$~\cite{supmat}.

\textit{Photon switch at finite-time resolution—}The recovery in Fig.~\ref{fig: lambda figures}(b) develops within a few cavity lifetimes, so observing it depends on how finely the detectors resolve the delay. We describe the detectors by a Gaussian error of rms width $s$ in the recorded delay; for two independent detectors with rms jitter $s_{\rm d}$ each, $s=\sqrt2\,s_{\rm d}$. Coincidences are counted in bins of width $w=0.5/\kappa$ centered at the delay $u$, and the two bins at zero delay and near the maximum of the rise are compared,
\begin{align}
 q_s(u)&=\int_{u-w/2}^{u+w/2}\!d\tau\int\!d\tau'\,\frac{e^{-(\tau-\tau')^2/2s^2}}{\sqrt{2\pi}\,s}\,P_{\rm VH}(\tau'),\nonumber\\
 C_s&=\frac{q_s(4/\kappa)}{q_s(0)}-1.
 \label{eq: bin contrast}
\end{align}

Here $q_s(u)$ is the mean number of coincidences per pulse in the bin. Near $u=0$, the jitter admits delays $\tau'<0$, at which the H photon is detected first. There, $P_{\rm VH}$ follows from Eq.~\eqref{eq: General G with noise} with the roles of the two detections exchanged and the pair operator $\hat B_{\rm VH}=\hat B_{\rm HV}$ evaluated at the H detection, as defined in the Supplemental Material.

Figure~\ref{fig: switch detector} shows $C_s$ for the pulse of Fig.~\ref{fig: lambda figures}. Without jitter, the full model gives $C_0=4.0\%$ and prodiabatic elimination $4.7\%$, whereas adiabatic elimination and the first line of Eq.~\eqref{eq: General G with noise} give $-2.3\%$ and $-2.1\%$. At $s=0.5/\kappa$ these values are $4.2\%$, $5.0\%$, $-1.2\%$ and $-0.9\%$. The sign of $C_s$ thus distinguishes the descriptions up to $s\simeq0.8/\kappa$, where the first line turns positive; adiabatic elimination follows at $s\simeq0.9/\kappa$. For the full model, $q_0(0)=2.64\times10^{-5}$ and $q_0(4/\kappa)=2.75\times10^{-5}$. With Poissonian counts, resolving their difference by five standard deviations requires $25[q_0(0)+q_0(4/\kappa)]/[q_0(4/\kappa)-q_0(0)]^2\simeq1.2\times10^9$ pulses, to be divided by the efficiencies $\eta_{\rm H}\eta_{\rm V}$ of the two detection channels.
}

\makeatletter
\newcommand{\seclabel}[2]{\phantomsection\protected@edef\@currentlabel{#2}\label{#1}}
\makeatother

\preprint{APS/123-QED}
\pagebreak
\widetext

\newpage
\begin{center}
\vskip0.5cm
{\Large Supplemental Material:\\
Prodiabatic Elimination:\\
Higher Order Elimination of Fast Variables with Quantum Noise}
\vskip0.1cm
{Jan Neuser$^{1,2}$, Marcelo Janovitch$^1$, Matteo Brunelli$^3$, and Patrick P. Potts$^1$}
\vskip0.1cm
{$^1$\textit{Department of Physics and Swiss Nanoscience Institute,
	\\ University of Basel, Klingelbergstrasse 82, 4056 Basel,
Switzerland }}
\vskip0.1cm
{$^2$ \textit{Atominstitut, TU Wien, 1020 Vienna, Austria}}
\vskip0.1cm
{$^3$\textit{JEIP, UAR 3573 CNRS, Coll\`ege de France, PSL Research University,\\
11 Place Marcelin Berthelot, 75321 Paris Cedex 05, France}}
\vskip0.5cm
{\today}
\vskip0.2cm
\vskip0.1cm
\end{center}
\vskip0.4cm

\setcounter{equation}{0}
\setcounter{figure}{0}
\setcounter{table}{0}
\setcounter{page}{1}
\renewcommand{\theequation}{S\arabic{equation}}
\renewcommand{\thefigure}{S\arabic{figure}}

\rev{This Supplement gives the vacuum-input model, the ordered-correlator derivation, the cavity-response spectrum, a compact Jaynes--Cummings benchmark, and the photon-switch calculation supporting the Letter.}

\section{1. Equivalence Between Lindblad Master Equation and Langevin Equation}\seclabel{sec: model}{1}
Equations~\eqref{eq: general dot a} and \eqref{eq: general dot b} can be obtained from the general quantum Langevin equation in the presence of two input channels,
\begin{align}
\dv{t}\hat{\mathcal{O}} =
i[\hat{H}, \hat{\mathcal{O}}]\label{eq:general_qle}
&- [\hat{\mathcal{O}}, \hat a^\dagger] \qty(\frac{\kappa}{2} \hat a +\sqrt{\kappa} \hat{a}_\text{in}) - \qty(\frac{\kappa}{2} \hat a^\dagger+\sqrt{\kappa} \hat{a}_\text{in}^\dagger) [\hat{\mathcal{O}},\hat a]\\
&- [\hat{\mathcal{O}}, \hat b^\dagger] \qty(\frac{\gamma}{2} \hat b +\sqrt{\gamma} \hat{b}_\text{in}) - \qty(\frac{\gamma}{2} \hat b^\dagger+\sqrt{\gamma} \hat{b}_\text{in}^\dagger) [\hat{\mathcal{O}},\hat b]\nonumber,
\end{align}
\rev{where the Hamiltonian is}
\begin{equation}
    \hat H = \hat  H_{\rm b} + \Delta \hat a^\dagger\hat  a + g\left(\hat a \hat b^\dagger + \hat  b \hat a^\dagger \right) - f(t) \left( \hat a+ \hat a^\dagger\right) \, ,
\end{equation}
and $\hat  H_{\rm b}$ models the atom in the absence of the cavity.  This is done
by setting the generic operator $\hat{\mathcal{O}} = \hat a,~\hat b$, and introducing $\Omega \hat v =  [\hat b, \hat H_{\rm b}]$.

In turn, Eq.~\eqref{eq:general_qle} is equivalent to the master equation~\cite{InputOutputTheory}
\begin{equation}
    \dot {\hat \rho} = - i [\hat H, \hat \rho ]+ \kappa \mathcal{D}[\hat a] \hat \rho + \gamma \mathcal{D}[\hat b ]\hat \rho \, ,
\end{equation}
which leads to the same equations of motion on the level of expectation values, such that for any operator $\hat{\mathcal{O}}$ its expectation value $\langle \dot{\hat{\mathcal{O}}}\rangle$ is the same for both equations.

\section{2. Derivation of the Prodiabatic Elimination}\seclabel{sec: derivation}{2}
The first step in deriving prodiabatic elimination is to expand the operator $\hat{b}(t-\tau) $ in the formal solution for $\hat{a}(t)$, see Eq.~\eqref{eq: formal solution to general poblem}. To this end, we write
\begin{equation}
    \hat{b}(t-\tau) = \hat{b}(t) - \int_0^\tau dt' \dot{\hat{b}}(t-t') \, .
    \label{eq:adbfirstint}
\end{equation}
We now rewrite Eq.~\eqref{eq: general dot b} as
\begin{equation}
\label{eq: dot b with X}
    \dot{\hat{b}}(t) = -\kappa\hat{X}(t)-\hat{r}(t) \sqrt{\gamma}\hat{b}_{\rm in}(t),
\end{equation}
with
\begin{equation}
\label{eq:xapp}
    \hat{X}(t)= \frac{i\Omega}{\kappa} \hat v(t) + \frac{\gamma}{2\kappa}  \hat r(t)\hat b(t) + \frac{ig}{\kappa}  \hat r(t)\hat a(t) \, .
\end{equation}
Inserting Eq.~\eqref{eq: dot b with X} into Eq.~\eqref{eq:adbfirstint} allows us to write
\begin{equation}
    \hat b(t-\tau) = \hat b(t)+\kappa \tau \hat X(t) + \frac{\sqrt{\gamma}}{\kappa}\int_0^\tau \kappa dt' \;\hat r(t-t')\hat b_{\rm in}(t-t') -\kappa \int_0^\tau dt' \; \int_0^{t'} dt'' \dot{\hat X}(t-t'')\,  .
    \label{eq:adbsecondint}
\end{equation}
To make progress, we note that $\hat{b} = \mathcal{O}(1)$ and we anticipate that $\hat{a}=\mathcal{O}(\epsilon)$ [see Eq.~\eqref{eq:lowestordera}]. As a consequence, $\hat{X}=\mathcal{O}(\epsilon^2)$. We further note that $\kappa\tau$ and $\kappa dt$ are $\mathcal{O}(1)$ due to the exponentials appearing in the integrals of Eq.~\eqref{eq: formal solution to general poblem} and \eqref{eq:adbsecondint}. Since the atom evolves on the timescale of $\epsilon$ (or slower) compared to the cavity, we have $\dot{\hat X}/\kappa=\mathcal{O}(\epsilon^3)$.

\subsection{2.1 Ignoring the Noise}\seclabel{sec: no noise}{2.1}
We first drop all contributions from noise terms $\hat{a}_{\rm in}$ and $\hat{b}_{\rm in}$. In the next section, we discuss how to properly include them. Ignoring the noise terms allows us to write
\begin{equation}
\label{eq:bnonoise}
    \hat b(t-\tau) = \hat b(t)+\kappa \tau \hat X(t)+\mathcal{O}(\epsilon^3).
\end{equation}
To lowest order in $\epsilon$, we may also drop the second term on the right-hand side. Inserting $\hat b(t-\tau)=\hat{b}(t)$ into Eq.~\eqref{eq: formal solution to general poblem} results in
\begin{equation}
    \label{eq:lowestordera}
    \hat{a} = -\frac{2 ig t_{\rm c}}{\kappa}\hat{b}(t)+i F(t)+\mathcal{O}(\epsilon^3).
\end{equation}
To include higher order terms, we insert Eq.~\eqref{eq:lowestordera} into the right-hand side of Eq.~\eqref{eq:xapp}. Inserting the resulting $\mathcal{O}(\epsilon^2)$ expression for $\hat{X}$ into Eq.~\eqref{eq:bnonoise}, and inserting that expression into Eq.~\eqref{eq: formal solution to general poblem} then results in $\hat{a}(t)=\hat{a}_{\rm pdb}(t)+\mathcal{O}(\epsilon^4)$, with $\hat {a}_{\rm pdb}$ given in Eq.~\eqref{eq: apdb without noise} in the Letter.

\subsection{2.2 Including the Noise}\seclabel{sec: noise}{2.2}

We now consider the terms that arise due to vacuum noise, starting with the third term on the right-hand side of Eq.~\eqref{eq:adbsecondint}. Inserting this term into Eq.~\eqref{eq: formal solution to general poblem} shows that this term contributes to the annihilation operator $\hat{a}$ as
\begin{equation}
    \rev{\hat{\xi}(t) = -ig\sqrt{\gamma}\int_0^\infty d\tau\,\chi_{\rm c}(\tau)\int_0^\tau dt'\,\hat r(t-t')\hat b_{\rm in}(t-t').}
\end{equation}
This operator then contributes to the expectation values of interest as
$ \langle \cdots \hat\xi(t) \hat a(t_k)\cdots \hat a(t_1)\rangle $. We now show that this contribution can be neglected. To this end, we consider
\begin{equation}
    [\hat{b}_{\rm in}(t-t'),\hat{a}(t_k)] = -\frac{2 ig t_{\rm c}}{\kappa}[\hat{b}_{\rm in}(t-t'),\hat{b}(t_k)] +\mathcal{O}(\epsilon^3),
    \label{eq:binacomm}
\end{equation}
where we used Eq.~\eqref{eq:lowestordera}.
This commutator is only finite when $t-t'\leq t_k$, as system operators can only be influenced by input operators from the past~\cite{InputOutputTheory}. Using Eq.~\eqref{eq:adbsecondint}, we may write
\begin{equation}
    \hat b(t_k) = \hat b(t-t'-\delta)-\kappa (t_k-t+t'+\delta) \hat X(t-t'-\delta) - \frac{\sqrt{\gamma}}{\kappa}\int_{t-t'-\delta}^{t_k} \kappa d\tau \hat r(\tau)\hat b_{\rm in}(\tau)+\mathcal{O}(\epsilon^3),
\end{equation}
where $\delta$ is an arbitrary small time. Inserting this expression into Eq.~\eqref{eq:binacomm}, we find
\begin{equation}
    [\hat{b}_{\rm in}(t-t'),\hat{a}(t_k)] = \mathcal{O}(\epsilon^3).
    \label{eq:binacomm2}
\end{equation}
As a consequence, we may commute the input noise operators appearing in $ \langle \cdots \hat\xi(t) \hat a(t_k)\cdots \hat a(t_1)\rangle $ all the way to the right, where they annihilate the vacuum. The correction to Eq.~\eqref{eq: sigma in time ordered expecvalue} arising from the $\epsilon^3$ term in Eq.~\eqref{eq:binacomm2} contributes to the order $\epsilon^{N+M+3}$ and can be neglected as our approximation is only valid to order $\epsilon^{N+M+2}$.

Next we consider the effect of the vacuum input noise of the cavity, which does contribute to Eq.~\eqref{eq: sigma in time ordered expecvalue} through the second and third line of Eq.~\eqref{eq: General G with noise}. From Eq.~\eqref{eq: formal solution to general poblem}, we can infer the contributions from the cavity noise to enter through terms of the form $ \langle \cdots \hat A_{\rm in}(t) \hat a(t_k)\cdots \hat a(t_1)\rangle $. We therefore consider
\begin{equation}
    [\hat{A}_{\rm in}(t),\hat{a}(t_k)] = i\frac{g}{\kappa}\rev{\int_0^\infty\kappa d\tau\int_0^\infty\kappa d\tau'} e^{-\frac{\kappa}{2t_{\rm c}}(\tau+\tau')}\frac{1}{\sqrt{\kappa}}[\hat{a}_{\rm in}(t-\tau),\hat b(t_k-\tau')],
    \label{eq:commaina}
\end{equation}
where we used Eq.~\eqref{eq: formal solution to general poblem}. Note that the right-hand side is only finite for $t-\tau\leq t_k-\tau'$. To make further progress, we rewrite Eq.~\eqref{eq:adbsecondint} as
\begin{equation}
     \hat b(t') = \hat b(t-\delta)-\kappa (t'-t+\delta) \hat X(t-\delta) - \int_{t-\delta}^{t'}\kappa d\tau\; \int_{t-\delta}^{\tau} \kappa d\tau' \dot{\hat X}(\tau')/\kappa\,  ,
     \label{eq:bapp3}
\end{equation}
where this time we dropped the noise term proportional to $\hat{b}_{\rm in}$ as it will not contribute. Equation \eqref{eq:bapp3} allows us to write
\begin{equation}
    [\hat{a}_{\rm in}(t),\hat{b}(t')] = -\int_{t-\delta}^{t'}\kappa d\tau\int_{t-\delta}^\tau \kappa d\tau'[\hat{a}_{\rm in}(t),\dot{\hat{X}}(\tau')/\kappa].
    \label{eq:commainb}
\end{equation}
To evaluate this expression, we need
\begin{equation}
    \label{eq:xdot}
    \dot{\hat{X}}(t) = \frac{g}{\kappa}\hat{a}^\dagger(t)\left[[\hat{r}(t),\hat{b}(t)]\left(g\hat{a}(t)-i\frac{\gamma}{2}\hat{b}(t)\right)+\Omega[\hat{v}(t),\hat{b}(t)]\right]+\cdots,
\end{equation}
where we only wrote out the terms that contribute to Eq.~\eqref{eq:commainb}.
Using
\begin{equation}
\label{eq:adag0}
    \hat{a}^\dagger(t) = -\sqrt{\kappa}\rev{(\chi_{\rm c}^**\hat{a}_{\rm in}^\dagger)(t)}+\mathcal{O}(\epsilon),
\end{equation}
we then find
\begin{equation}
    \label{eq:commainxdot}
    \frac{1}{\sqrt{\kappa}}[\hat{a}_{\rm in}(t),\dot{\hat{X}}(\tau')/\kappa] = -\frac{g}{\kappa^2}\rev{\chi_{\rm c}^*(\tau'-t)}\left[[\hat{r}(t),\hat{b}(t)]\left(g\hat{a}(t)-i\frac{\gamma}{2}\hat{b}(t)\right)+\Omega[\hat{v}(t),\hat{b}(t)]\right]+\mathcal{O}(\epsilon^4),
\end{equation}
for $t\leq \tau'$ and zero otherwise. Inserting Eq.~\eqref{eq:commainxdot} into Eq.~\eqref{eq:commainb}, and inserting that equation into Eq.~\eqref{eq:commaina} we find
\begin{equation}
    [\hat{A}_{\rm in}(t),\hat{a}(t')] = \rev{\chi_{\rm c}(t-t')}\hat{B}(t')+ \mathcal{O}\left(\epsilon^5 \right),
    \label{eq:commainafinal}
\end{equation}
for $t\geq t'$, where $\hat{B}(t)$ is given in Eq.~\eqref{eq:Bop} in the Letter.

Knowing how the noise terms contribute to the correlation functions in Eq.~\eqref{eq: sigma in time ordered expecvalue}, we can find the appropriate expansion of $\hat{a}$ by approximating Eq.~\eqref{eq:adbsecondint} with Eqs.~\eqref{eq:xdot} and \eqref{eq:adag0}, dropping terms including $\hat{b}_{\rm in}$, and inserting back into the formal solution in Eq.~\eqref{eq: formal solution to general poblem}. This results in Eq.~\eqref{eq: general a_pdb with noise} in the Letter, which respects Eq.~\eqref{eq:commainafinal}. To find the correlation functions in Eq.~\eqref{eq: sigma in time ordered expecvalue} up to order $\epsilon^{N+M+2}$, we insert the expansion of $\hat{a}$ in Eq.~\eqref{eq: general a_pdb with noise} into Eq.~\eqref{eq: sigma in time ordered expecvalue} and use the commutation relation in Eq.~\eqref{eq:commainafinal} repeatedly, only keeping terms up to the desired order. This results in our main result, Eq.~\eqref{eq: General G with noise} in the Letter.

{\color{blue}
\section{3. Correlation Spectrum and Cavity Response}\seclabel{sec: spectrum}{3}
\textit{Detection convention.} In a vacuum-input channel, normally and time-ordered output correlations reduce to those of $\sqrt{\kappa_{\rm d}}\hat a$, where $\kappa_{\rm d}$ is the monitored port's loss rate. The total linewidth is $\kappa$. An equivalent measurement displaces the coherent drive field before detection. For reflection from the driven port, the drive amplitude must instead be included, as in Sec.~\ref{sec: switch}. Photocount factorial moments are integrals of these ordered output correlations~\cite{KelleyKleiner1964,MandelWolf1995}. For detected mean flux $R_{\rm d}=\eta\kappa_{\rm d}\langle\hat a^\dagger\hat a\rangle$, the ideal two-sided point-process spectrum is
\begin{equation}
 S_I(\omega)=R_{\rm d}+R_{\rm d}^2S(\omega).
 \label{eq: photocurrent spectrum}
\end{equation}
The first term is shot noise; the filter law concerns the excess spectrum $S$, which has units of time. Detector jitter smears time-resolved correlations but preserves integrated counts away from window boundaries; dead time and spectral filtering require their own measurement models.

\textit{Fast response and initial slope.} Set $n=\langle\hat a^\dagger\hat a\rangle$ and $\lambda=\kappa/(2t_{\rm c})$. For $N=M=2$ in Eq.~\eqref{eq: General G with noise}, the pair terms are complex conjugates. At leading order on the cavity timescale, atomic operators can be evaluated at equal times, giving
\begin{equation}
 g^{(2)}_{\rm pair}(t)=-\Re\left[\frac{e^{-\lambda t}}\lambda\mathcal C\right],\qquad
 \mathcal C=\frac{\langle\hat a_{\rm adb}^{\dagger2}[\hat a_{\rm adb},\dot{\hat a}_{\rm adb}]\rangle}{\langle\hat a_{\rm adb}^\dagger\hat a_{\rm adb}\rangle^2}.
 \label{eq: SM noise part}
\end{equation}
Let $s(t)$ denote the slow contribution to $g^{(2)}(t)-1$. Stationarity of its equal-time fourth moment gives $\Re\mathcal C=-s'(0)$ to leading order. Thus the pair term cancels the leading initial slope. For the full stationary model of Sec.~\ref{sec: model}, quantum regression and the adjoint master equation give,
\begin{equation}
 \left.\frac{dg^{(2)}}{dt}\right|_{0^+}
 =\frac{-\kappa\langle\hat a^{\dagger2}\hat a^2\rangle+2\Re\langle\hat a^{\dagger2}\hat a(-ig\hat b+if)\rangle}{n^2}=0.
 \label{eq: SM general flat top}
\end{equation}
The numerator vanishes due to the earlier outlined stationarity of $\langle\hat a^{\dagger2}\hat a^2\rangle$. This exact zero-slope identity holds at finite drive and detuning for the specified model, something that cannot be captured by adiabatic elimination, as it disregards the pair term. It describes smooth zero-delay behavior; it does not imply a dead time or that the stationary point is a maximum.

\textit{Filter law.} At frequencies above all slow dynamical frequencies and decay rates, integration by parts gives the leading slow tail $-2s'(0)/\omega^2$. Adding the transform of Eq.~\eqref{eq: SM noise part} yields
\begin{equation}
 S(\omega)\simeq2\Re\left[\mathcal C\left(\frac1{\omega^2}-\frac1{\lambda^2+\omega^2}\right)\right]
 =\Re\left[\frac{\lambda^2}{\lambda^2+\omega^2}\frac{2\mathcal C}{\omega^2}\right].
 \label{eq: SM general spectrum}
\end{equation}
On cavity resonance, $\lambda=\kappa/2$ is real and the filter factors out, giving Eq.~\eqref{eq: filter law}. Off resonance, the complex coefficient $\mathcal C$ matters; there is no universal real multiplier of the measured $S_{\rm adb}$. The relative resonant law requires a nonzero leading tail and is taken at fixed $\omega/\kappa=\mathcal O(1)$. It is nonuniform at cancellations, such as $F_{\rm p}=1$ in the resonant weak-drive JC model, and does not establish unrestricted high-frequency accuracy.

\textit{Integrated observables.} For a stationary correlation with a nonzero slow integral, integrating a pair contribution over a window long compared with the atomic relaxation time suppresses its relative weight by another factor of the slow-to-cavity rate ratio. It then enters at relative order $\epsilon^4$, alongside omitted amplitude corrections. Adding this term alone need not improve the integrated result. This order estimate does not apply uniformly near a vanishing slow integral or to arbitrary pulsed protocols.
}

{\color{blue}
\section{4. Jaynes--Cummings Benchmark}\seclabel{sec: JC}{4}
The constant-drive full model used for the numerical comparisons is
\begin{align}
 \dot{\hat\rho}&=-i[\hat H,\hat\rho]+\kappa\mathcal D[\hat a]\hat\rho+\gamma\mathcal D[\hat\sigma]\hat\rho,\label{eq: Lindblad Master Equation System}\\
 \hat H&=\Delta\hat a^\dagger\hat a+\frac\Omega2\hat\sigma_z+g(\hat a^\dagger\hat\sigma+\hat\sigma^\dagger\hat a)-f(\hat a+\hat a^\dagger),\label{eq: System Hamiltonian}
\end{align}
with $\mathcal D[\hat L]\hat\rho=\hat L\hat\rho\hat L^\dagger-\{\hat L^\dagger\hat L,\hat\rho\}/2$. The ground-state atom and vacuum cavity initialize the population comparison; the correlation benchmarks use the stationary state.
\subsection{4.1 Model and Reduced Generator}\seclabel{sec: JC model}{4.1}

The reduced atomic equations, Eqs.~\eqref{eq:avgjc1} and \eqref{eq:avgjc2}, are reproduced to the retained order by
\begin{equation}
        \dot{\hat \rho} = - i [ \hat{H}_\text{pdb},\hat \rho] + \Gamma_1  \mathcal{D}[\hat \sigma] \hat\rho
    + \Gamma_0 \mathcal D[\hat \sigma+ \xi \hat\sigma_z]\hat\rho \, ,
    \label{eq: Lme 2l system supp}
\end{equation}
where
\begin{equation}
    \hat H_{\rm pdb} = c_z \hat \sigma_z + c_x \hat \sigma_x + c_y \hat \sigma_y
\end{equation}
with $\hat\sigma_z = [\hat\sigma^\dagger, \hat\sigma]$, $\hat\sigma_x = \hat\sigma + \hat\sigma^\dagger$ and $\hat{\sigma}_y = i(\hat{\sigma}-\hat{\sigma}^\dagger)$. Further, the coefficients read
\begin{align}
    c_z  =& \frac{\Omega}{2} + \frac{\gamma}{2\kappa}F_{\rm p} \left(\Omega \Re{t_{\rm c}^2}-\Delta \abs{t_{\rm c}}^2\right)- \Delta \abs{t_{\rm c}}^4\frac{\gamma^2}{\kappa^2}F_{\rm p}\left[ F_{\rm p} \left(2\abs{t_{\rm c}}^2-\frac{1}{2}\right)+1\right]\\
    c_x  =& \frac{2fg\Delta}{\kappa^2} \abs{t_{\rm c}}^2\left[ \frac{\gamma}{\kappa} F_{\rm p} \left( \Re{t_{\rm c}^2}+ 2 \abs{t_{\rm c}}^4\right) + 2\right]\\
    c_y  =& - \frac{2gf}{\kappa}\abs{t_{\rm c}}^2\left[ \frac{\gamma}{\kappa}F_{\rm p}\left( \frac{3}{2}\Re{t_{\rm c}^2}- \abs{t_{\rm c}}^4\right)+1\right] \, .
\end{align}
The rates in Eq.~\eqref{eq: Lme 2l system supp} are given by
\begin{align}
    \Gamma_0  =& \gamma(1+ F_{\rm p}\abs{t_{\rm c}}^2)\\
    \Gamma_1  =& \frac{8\Delta \Omega}{\kappa}\frac{\gamma}{\kappa}F_{\rm p} \abs{t_{\rm c}}^4  + \gamma \frac{\gamma}{\kappa}F_{\rm p}\abs{t_{\rm c}}^2\left\{ 2 \abs{t_{\rm c}}^2\left[F_{\rm p}\left(2\abs{t_{\rm c}}^2- \frac{3}{2}\right)+1\right]-1\right\} \\
    \xi  =& \frac{2f g t_{\rm c}^3}{\kappa^2}\frac{F_{\rm p}}{1+F_{\rm p}\abs{t_{\rm c}}^2} \, .
\end{align}
Equation \eqref{eq: Lme 2l system supp} will reproduce the equations of motion in Eqs.~\eqref{eq:avgjc1} and \eqref{eq:avgjc2}, up to terms that are proportional to $\gamma \epsilon^4$, which can safely be neglected. \rev{The displayed dissipator representation is not guaranteed to be completely positive at arbitrary detuning: $\Gamma_1$ can be negative. It reproduces the retained perturbative equations, while a negative eigenvalue of its dissipative coefficient matrix can occur beyond that order. At resonance both displayed rates are nonnegative.} At resonance, $\Gamma_0=\Gamma_{\rm adb}$, $\Gamma_1=\Gamma_{\rm adb}\gamma F_{\rm p}/\kappa$, $c_z=c_x=0$ and $c_y=-(2gf/\kappa)(1+\gamma F_{\rm p}/2\kappa)$.

Regression~\cite{Quantumregressiontheoremformulti-timecorrelators,QuantumMasterEQ_Breuer} with this generator and the reconstructed observable gives the first line of Eq.~\eqref{eq: General G with noise} to the retained order. Its atomic rates are slow; it does not supply the separate pair terms decaying at $\kappa/2$. This statement concerns a time-independent slow generator and instantaneous observable maps, not reduced descriptions supplemented by a memory kernel or a detection-conditioned reconstruction.

\subsection{4.2 Exact Weak-Drive Correlation and Error Order}\seclabel{sec: JC exact}{4.2}
To benchmark the reduced correlation, we calculate $g^{(2)}$ analytically at leading order in the drive.
For the resonant Jaynes--Cummings model, a cavity detection prepares $|\phi\rangle=\hat a|\psi_{\rm ss}\rangle$. Writing $\alpha=\langle g,1|\phi\rangle$ and $\beta=\langle e,0|\phi\rangle$, the conditional one-excitation amplitudes obey
\begin{equation}
 \frac{d}{dt}\begin{pmatrix}\alpha\\\beta\end{pmatrix}
 =\begin{pmatrix}-\kappa/2&-ig\\-ig&-\gamma/2\end{pmatrix}
 \begin{pmatrix}\alpha\\\beta\end{pmatrix}
 +\begin{pmatrix}ifc_{g1}\\0\end{pmatrix},
 \qquad c_{g1}=\langle g,1|\psi_{\rm ss}\rangle=\frac{2if}{\kappa(1+F_{\rm p})}.
 \label{eq: SM conditional amplitudes}
\end{equation}
The source is the surviving $|g,0\rangle$ amplitude after detection. Define
\begin{equation}
 A=1+\frac{\gamma(1+F_{\rm p})}{\kappa},\qquad
 R=\sqrt{(\kappa-\gamma)^2-16g^2},\qquad
 \Gamma=\frac{\kappa+\gamma-R}{2},\quad
 \Gamma_{\rm f}=\frac{\kappa+\gamma+R}{2}.
 \label{eq: SM two rates}
\end{equation}
For the overdamped regime $16g^2<(\kappa-\gamma)^2$, the conditional amplitudes relax at the rates $\Gamma/2$ and $\Gamma_{\rm f}/2$, and the weak-drive correlation is 
\begin{equation}
 g^{(2)}(t)=[1-De^{-\Gamma t/2}+Ee^{-\Gamma_{\rm f}t/2}]^2,\qquad
 D=\frac{F_{\rm p}^2\Gamma_{\rm f}}{AR},\quad E=\frac{F_{\rm p}^2\Gamma}{AR}.
 \label{eq: SM exact g2}
\end{equation}
This is exact in $F_{\rm p}$ and $\gamma/\kappa$ at leading order in drive. At fixed $F_{\rm p}$, $\Gamma=\Gamma_{\rm pdb}[1+\mathcal O((\gamma/\kappa)^2)]$, $D=F_{\rm p}^2[1+\mathcal O((\gamma/\kappa)^2)]$, and $E=(\gamma/\kappa)F_{\rm p}^2(1+F_{\rm p})+\mathcal O((\gamma/\kappa)^2)$. The last amplitude produces the pair correction in Eq.~\eqref{eq: g2 first order in f}; the first correction to $D$ lies beyond its controlled order. Also $D\Gamma=E\Gamma_{\rm f}$, giving the exact zero-slope identity of Sec.~\ref{sec: spectrum}. The prodiabatic formula cancels the slope only to leading order.

At zero delay, writing $x=\Gamma_{\rm adb}/\kappa$,
\begin{equation}
 g^{(2)}(0)=\left(1-\frac{F_{\rm p}^2}{1+x}\right)^2
 =(1-F_{\rm p}^2)^2+2F_{\rm p}^2(1-F_{\rm p}^2)x+F_{\rm p}^2(3F_{\rm p}^2-2)x^2+\mathcal O(x^3).
 \label{eq: SM g20}
\end{equation}
Adiabatic elimination keeps the first term; prodiabatic elimination also keeps the second. Their leading absolute relative errors away from zeros are therefore $2F_{\rm p}^2x/|1-F_{\rm p}^2|$ and $|F_{\rm p}^2(3F_{\rm p}^2-2)|x^2/(1-F_{\rm p}^2)^2$. Near a vanishing leading correlation the relative expansion is nonuniform, and a truncated formula can become negative at an uncontrolled order.

A positive-delay weak-drive zero solves $1-De^{-\Gamma t/2}+Ee^{-\Gamma_{\rm f}t/2}=0$ and, in this overdamped regime, exists for $F_{\rm p}>1+\gamma/\kappa$. The estimate $t_*\simeq2\ln D/\Gamma$ neglects the fast exponential. At the parameters of Fig.~\ref{fig: three figures}(a,b), it gives $\kappa t_*\simeq110.10$; the finite-drive full-model minimum is at $\kappa t\simeq110.11$ with $g^{(2)}\simeq0.00413$. Adiabatic and prodiabatic weak-drive formulas give $121.70$ and $111.65$. Their difference is due to the corrected slow rate.

\subsection{4.3 Spectrum and Numerical Checks}\seclabel{sec: JC checks}{4.3}
Transforming each exponential in Eq.~\eqref{eq: SM exact g2} gives
\begin{equation}
 S(\omega)=\frac{-2D\Gamma}{(\Gamma/2)^2+\omega^2}
 +\frac{2D^2\Gamma}{\Gamma^2+\omega^2}
 +\frac{2E\Gamma_{\rm f}}{(\Gamma_{\rm f}/2)^2+\omega^2}
 +\frac{2E^2\Gamma_{\rm f}}{\Gamma_{\rm f}^2+\omega^2}
 -\frac{2DE(\kappa+\gamma)}{[(\kappa+\gamma)/2]^2+\omega^2}.
 \label{eq: SM spectrum}
\end{equation}
The last three terms are the fast contribution and can subtract spectral weight. Their leading cancellation with the slow tail gives the resonant filter derived in Sec.~\ref{sec: spectrum}. No absolute values are taken in the spectral ratios of Fig.~\ref{fig: three figures}(c).

The full stationary master equation and its resolvent benchmark the weak-drive expressions.
Figure~\ref{fig: three figures}(a,b) uses $f/\kappa=2.5\times10^{-4}$; Fig.~\ref{fig: three figures}(c,d) uses Eq.~\eqref{eq: SM exact g2} and Eq.~\eqref{eq: SM spectrum},
checked against the full model as the drive is reduced.

\begin{figure}[htb]
 \centering
 \includegraphics[width=0.62\linewidth]{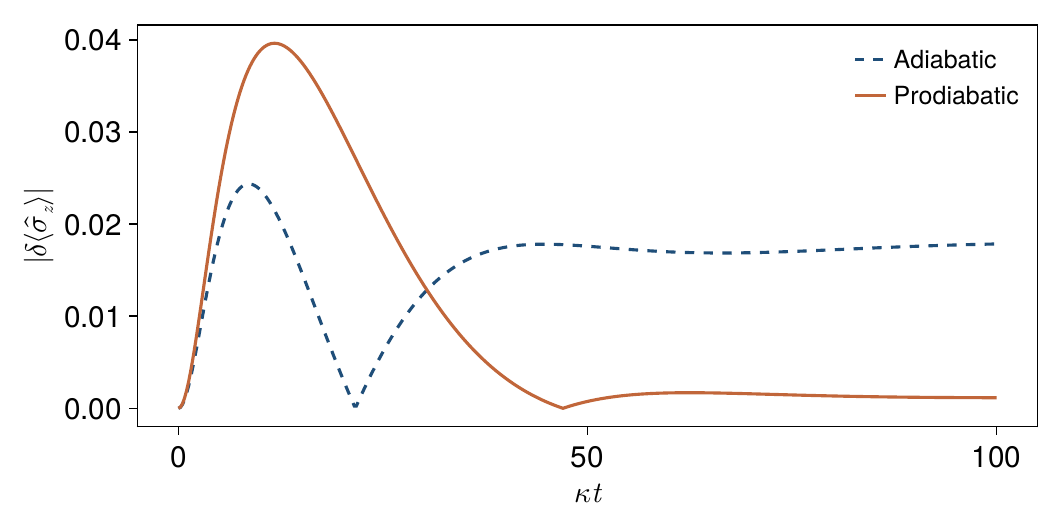}
 \caption{\rev{Absolute atomic-population errors for $g/\kappa=0.15$, $\gamma/\kappa=0.005$, $\Delta/\kappa=0.05$, $\Omega/\kappa=0.0005$ and $f/\kappa=0.1$. The atom starts in its ground state and the full cavity in vacuum. Blue dashed: adiabatic; orange: prodiabatic. The reduced constant-drive response is present immediately, whereas the full cavity builds up from vacuum.}}
 \label{fig: population check}
\end{figure}
For the 401-point grid $0\leq\kappa t\leq100$, the adiabatic/prodiabatic maximum errors are $0.0244/0.0397$ and the RMS errors $0.0163/0.0160$. The endpoint errors at $\kappa t=100$ are $0.0179/0.00115$, a reduction by a factor $15.5$. This endpoint gain is not a uniform transient improvement. The far-past initial condition of Eq.~\eqref{eq: formal solution to general poblem} excludes an arbitrary initial cavity transient.
}

{\color{blue}
\section{5. \NoCaseChange{Photon Switch}}\seclabel{sec: switch}{5}

We derive the conditional H field, Eq.~\eqref{eq: conditional field}, the coincidence density and its closed-switch reference, and the conditional reflectance. We distinguish the prodiabatic reconstruction, the weak-drive response of the atom, and the broad-pulse limits used in the Letter and its End Matter.

\subsection{5.1 Model and Reduced Dynamics}\seclabel{sec: switch model}{5.1}

With $\hat\sigma_{ij}=\ket{i}\!\bra{j}$, the resonant model is
\begin{align}
 \dot{\hat\rho}&=-i[\hat H,\hat\rho]
 +\kappa\sum_{\mu\in\{\mathrm{H,V}\}}\mathcal D[\hat a_\mu]\hat\rho
 +\gamma\sum_{j=1,2}\mathcal D[\hat\sigma_{j3}]\hat\rho,\label{eq: SM switch master}\\
 \hat H&=g(\hat a_{\rm H}^\dagger\hat\sigma_{13}
              +\hat a_{\rm V}^\dagger\hat\sigma_{23}+\mathrm{H.c.})
              -f(t)(\hat a_{\rm H}+\hat a_{\rm H}^\dagger),\label{eq: SM switch Hamiltonian}
\end{align}
with both modes in vacuum and the atom in $\ket{1}$ initially. We take a real pulse $f(t)$ carrying $\bar n=\int dt\,f(t)^2/\kappa$ incident photons. The Gaussian pulse $f(t)=f_0e^{-(t-t_0)^2/(2\sigma^2)}$ is used for Fig.~\ref{fig: lambda figures} and for the explicit limits below. Equations~\eqref{eq: general dot a} and \eqref{eq: general dot b} hold for $\hat a=(\hat a_{\rm H},\hat a_{\rm V})$ and $\hat b=(\hat b_{\rm H},\hat b_{\rm V})=(\hat\sigma_{13},\hat\sigma_{23})$, with
\begin{equation}
 \hat r_{\mu\nu}=[\hat b_\mu,\hat b_\nu^\dagger]=\hat\sigma_{\mu\nu}-\delta_{\mu\nu}\hat\sigma_{33},\qquad
 [\hat b_\ell,\hat r_{\mu\nu}]=-\delta_{\nu\ell}\hat b_\mu-\delta_{\mu\nu}\hat b_\ell,
 \label{eq: SM channel commutator}
\end{equation}
where the atomic indices H and V stand for 1 and 2. Since $\sum_\nu\hat r_{\mu\nu}\hat b_\nu=2\hat b_\mu$, Eq.~\eqref{eq: apdb without noise} gives
\begin{align}
 \hat a_{\rm pdb,H}&=-\frac{2ig}{\kappa}\Big(1+\frac{2\Gamma_{\rm adb}}{\kappa}\Big)\hat\sigma_{13}
 +iF+i\frac{\gamma F_{\rm p}}{\kappa}F\,(\hat\sigma_{11}-\hat\sigma_{33}),\label{eq: SM reconstructed H}\\
 \hat a_{\rm pdb,V}&=-\frac{2ig}{\kappa}\Big(1+\frac{2\Gamma_{\rm adb}}{\kappa}\Big)\hat\sigma_{23}
 +i\frac{\gamma F_{\rm p}}{\kappa}F\,\hat\sigma_{21}.
 \label{eq: SM reconstructed V}
\end{align}
Inserting these into the Heisenberg equations of the atom, with the products expanded to the order of Eq.~\eqref{eq: General G with noise}, closes the equations for the populations and the real coherence $\langle\hat\sigma_{13}\rangle$,
\begin{align}
 \langle\dot{\hat\sigma}_{11}\rangle&=\Gamma_{\rm pdb}\langle\hat\sigma_{33}\rangle-2gF\Big(1+\frac{\gamma F_{\rm p}}{\kappa}\Big)\langle\hat\sigma_{13}\rangle,\nonumber\\
 \langle\dot{\hat\sigma}_{22}\rangle&=\Gamma_{\rm pdb}\langle\hat\sigma_{33}\rangle-2gF\,\frac{\gamma F_{\rm p}}{\kappa}\langle\hat\sigma_{13}\rangle,\label{eq: SM reduced switch equations}\\
 \langle\dot{\hat\sigma}_{13}\rangle&=-\Gamma_{\rm pdb}\langle\hat\sigma_{13}\rangle
 +gF\Big[\Big(1+\frac{2\gamma F_{\rm p}}{\kappa}\Big)\langle\hat\sigma_{11}\rangle-\Big(1-\frac{\gamma F_{\rm p}}{\kappa}\Big)\langle\hat\sigma_{33}\rangle\Big],\nonumber
\end{align}
with $\langle\hat\sigma_{33}\rangle=1-\langle\hat\sigma_{11}\rangle-\langle\hat\sigma_{22}\rangle$, where
\begin{equation}
 \Gamma_{\rm pdb}=\Gamma_{\rm adb}\Big(1+\frac{2\gamma F_{\rm p}}{\kappa}\Big)
 \label{eq: SM pdb rate}
\end{equation}
is the atomic coherence decay rate returned by prodiabatic elimination, the counterpart of $\Gamma_{\rm adb}=\gamma(1+F_{\rm p})$. Adiabatic elimination drops all terms proportional to $\gamma F_{\rm p}/\kappa$ and $\Gamma_{\rm adb}/\kappa$ and replaces $F$ by $2f/\kappa$.

Following Sec.~\ref{sec: noise} with mode indices, the commutator below Eq.~\eqref{eq: response kernel} becomes $[\hat A_{{\rm in},\mu}(t'),\hat a_\nu(t)]=\chi_{\rm c}(t'-t)\hat B_{\mu\nu}(t)$ for $t'\geq t$, with $\hat B_{\mu\nu}=(4g^2/\kappa^3)[\hat b_\nu,\dot{\hat b}_{{\rm adb},\mu}]$ and $\dot{\hat b}_{{\rm adb},\mu}=-\Gamma_{\rm adb}\hat b_\mu+gF\sum_\ell\hat r_{\mu\ell}\delta_{\ell{\rm H}}$. Since $\hat b_\mu\hat b_\nu=0$, only the drive contributes, and Eq.~\eqref{eq: SM channel commutator} gives
\begin{equation}
 \hat B_{\mu\nu}=-\frac{4g^3F}{\kappa^3}\left(\delta_{\nu{\rm H}}\hat b_\mu+\delta_{\mu{\rm H}}\hat b_\nu\right),
 \label{eq: SM switch pair operator}
\end{equation}
whose off-diagonal element is $\hat B_{\rm HV}=-4g^3F\hat\sigma_{23}/\kappa^3$. In Eq.~\eqref{eq: General G with noise}, $\chi_{\rm c}(t_j-t_i)\hat B_{\mu\nu}(t_i)$ replaces the pair $\hat a_\mu(t_j)\hat a_\nu(t_i)$. The coherent part of the detected field $\hat c_{\rm H}$ commutes with these operators and leaves $\hat B_{\mu\nu}$ unchanged.

\subsection{5.2 Conditional H Field After a Raman Detection}\seclabel{sec: switch field}{5.2}

\textit{Exact.} Starting from $\ket{1}$, every state with a V photon has the atom in $\ket{2}$, so a V detection at time $t$ leaves the atom in $\ket{2}$ and the V mode empty, for any drive strength. From then on the atom is decoupled, and Eq.~\eqref{eq: SM switch master} gives, for the averages conditioned on the detection and $\tau\geq0$,
\begin{align}
 \langle\hat a_{\rm H}(t+\tau)\rangle_{\rm V}&=iF(t+\tau)+\chi_{\rm c}(\tau)\big[\langle\hat a_{\rm H}(t)\rangle_{\rm V}-iF(t)\big],\label{eq: SM conditional solution}\\
 \langle\hat c_{\rm H}^\dagger\hat c_{\rm H}(t+\tau)\rangle_{\rm V}&=\kappa\Big|\langle\hat a_{\rm H}(t+\tau)\rangle_{\rm V}-\frac{if(t+\tau)}{\kappa}\Big|^2
 +\kappa\chi_{\rm c}(\tau)^2\big[\langle\hat a_{\rm H}^\dagger\hat a_{\rm H}(t)\rangle_{\rm V}-|\langle\hat a_{\rm H}(t)\rangle_{\rm V}|^2\big].
 \label{eq: SM conditional intensity}
\end{align}
The H field thus relaxes to the empty-cavity response $iF$ from the value it had at the Raman detection. The coincidence density is $P_{\rm VH}(\tau)=\int dt\,\langle\hat c_{\rm V}^\dagger(t)\hat c_{\rm H}^\dagger(t+\tau)\hat c_{\rm H}(t+\tau)\hat c_{\rm V}(t)\rangle$, as defined in Eq.~\eqref{eq: switch profile} of the Letter. Multiplying Eq.~\eqref{eq: SM conditional intensity} by the Raman rate $\kappa\langle\hat a_{\rm V}^\dagger\hat a_{\rm V}\rangle(t)$ and integrating over $t$ gives this density exactly. The variance term contributes $1.3\times10^{-7}$ of $P_{\rm VH}(0)$ at the parameters of Fig.~\ref{fig: lambda figures}, so the profile is set by the conditional mean field.

\textit{Prodiabatic elimination.} Equation~\eqref{eq: General G with noise} with $N=2$, $M=1$ gives, for $t'=t+\tau$,
\begin{equation}
 \langle\hat a_{\rm V}^\dagger(t)\hat a_{\rm H}(t')\hat a_{\rm V}(t)\rangle
 =\big\langle\hat a_{\rm pdb,V}^\dagger(t)\,\hat a_{\rm pdb,H}(t')\,\hat a_{\rm pdb,V}(t)\big\rangle
 +\chi_{\rm c}(\tau)\big\langle\hat a_{\rm pdb,V}^\dagger(t)\hat B_{\rm HV}(t)\big\rangle.
 \label{eq: SM conditional pdb}
\end{equation}
In the first term, the quantum regression theorem propagates $\hat a_{\rm pdb,V}\hat\rho\,\hat a_{\rm pdb,V}^\dagger$, which is proportional to $\ket{2}\!\bra{2}$ and therefore stationary. Since $\hat a_{\rm pdb,H}\ket{2}=iF\ket{2}$, the first term equals $iF(t')\langle\hat a_{\rm V}^\dagger\hat a_{\rm V}\rangle(t)$: the first line of Eq.~\eqref{eq: General G with noise} returns the empty-cavity response and nothing else. The second term is $\chi_{\rm c}(\tau)(2ig/\kappa)(-4g^3F/\kappa^3)\langle\hat\sigma_{33}\rangle$. Dividing by $\langle\hat a_{\rm V}^\dagger\hat a_{\rm V}\rangle=(4g^2/\kappa^2)\langle\hat\sigma_{33}\rangle$ to leading order gives Eq.~\eqref{eq: conditional field}; at $\tau=0$ it fixes the bracket in Eq.~\eqref{eq: SM conditional solution},
\begin{equation}
 \langle\hat a_{\rm H}(t)\rangle_{\rm V}-iF(t)=-i\,\frac{\gamma F_{\rm p}}{2\kappa}F(t).\label{eq: SM conditional displacement}
\end{equation}
The exact displacement agrees with Eq.~\eqref{eq: SM conditional displacement} up to a relative correction of order $\gamma F_{\rm p}/\kappa$, which Eq.~\eqref{eq: General G with noise} does not control. 

\subsection{5.3 Coincidence Density}\seclabel{sec: switch coincidences}{5.3}

The coincidence density is the Raman rate times the conditional H intensity, integrated over the herald time,
\begin{equation}
 P_{\rm VH}(\tau)=\int\!dt\,\langle\hat c_{\rm V}^\dagger\hat c_{\rm V}(t)\rangle
 \langle\hat c_{\rm H}^\dagger\hat c_{\rm H}(t+\tau)\rangle_{\rm V},\qquad \tau\geq0.
 \label{eq: SM coincidence definition}
\end{equation}
This is the conditional-average identity used in Eq.~\eqref{eq: switch profile} of the End Matter; it does not require a weak drive. Equation~\eqref{eq: General G with noise} with $N=M=2$ and the detected fields $\hat c_\mu$ gives, 
by the argument below Eq.~\eqref{eq: SM conditional pdb},

\begin{equation}
 G^{(2)}_{\rm VH}(t,t+\tau)=\kappa^2\Big[F(t+\tau)-\frac{f(t+\tau)}{\kappa}\Big]^2\langle\hat a_{\rm V}^\dagger\hat a_{\rm V}\rangle(t)
 -(\gamma F_{\rm p})^2\chi_{\rm c}(\tau)F(t)\Big[F(t+\tau)-\frac{f(t+\tau)}{\kappa}\Big]\langle\hat\sigma_{33}(t)\rangle,
 \label{eq: SM switch G2}
\end{equation}
where the second term is the sum of the two pair lines, which are equal. This is the Raman rate times the conditional reflected intensity, as quoted in the Letter, with Eq.~\eqref{eq: conditional field} inserted and the square expanded to first order in the pair term. Expanding the Raman rate to the same order with Eq.~\eqref{eq: SM reconstructed V},
\begin{equation}
 \kappa\langle\hat a_{\rm V}^\dagger\hat a_{\rm V}\rangle=\gamma F_{\rm p}\Big[\Big(1+\frac{4\Gamma_{\rm adb}}{\kappa}\Big)\langle\hat\sigma_{33}\rangle-\frac{4gF}{\kappa}\,\mathrm{Re}\langle\hat\sigma_{13}\rangle\Big],
 \label{eq: SM raman rate}
\end{equation}
we obtain the explicit prodiabatic coincidence density
\begin{equation}
 \begin{aligned}
 P_{\rm VH}^{\rm pdb}(\tau)=\gamma F_{\rm p}\!\int\!dt\,\Big[F(t+\tau)-\frac{f(t+\tau)}{\kappa}\Big]\Big\{&
 \kappa\Big[F(t+\tau)-\frac{f(t+\tau)}{\kappa}\Big]
 \Big[\Big(1+\frac{4\Gamma_{\rm adb}}{\kappa}\Big)\langle\hat\sigma_{33}(t)\rangle-\frac{4gF(t)}{\kappa}\langle\hat\sigma_{13}(t)\rangle\Big]\\
 &-\gamma F_{\rm p}\,\chi_{\rm c}(\tau)F(t)\,\langle\hat\sigma_{33}(t)\rangle\Big\},
 \end{aligned}
 \label{eq: SM switch pdb}
\end{equation}
which contains only the parameters of Eqs.~\eqref{eq: SM switch master} and \eqref{eq: SM switch Hamiltonian}, the drive $f$, its filtered form $F=\chi_{\rm c}*f$, and the two atomic averages of Eq.~\eqref{eq: SM reduced switch equations}. Adiabatic elimination has $F=2f/\kappa$, no pair term and $\kappa\langle\hat a_{\rm V}^\dagger\hat a_{\rm V}\rangle=\gamma F_{\rm p}\langle\hat\sigma_{33}\rangle$, which yields the adiabatic coincidence density $P_{\rm VH}^{\rm adb}(\tau)=(\gamma F_{\rm p}/\kappa)\int dt\,\langle\hat\sigma_{33}(t)\rangle f(t+\tau)^2$. Retaining the corrected operators and dynamics but dropping the pair term defines the curve labeled first line in Fig.~\ref{fig: lambda figures}.

These expressions retain the order of Eq.~\eqref{eq: General G with noise}. We obtain the atomic averages by solving Eq.~\eqref{eq: SM reduced switch equations}. The closed-switch reference uses the same Raman rate but the empty-cavity H intensity,
\begin{equation}
 \begin{aligned}
 P_{\rm VH}^{\rm cl}(\tau)
 &=\int\!dt\,\kappa\langle\hat a_{\rm V}^\dagger\hat a_{\rm V}\rangle(t)\,
 \kappa\Big[F(t+\tau)-\frac{f(t+\tau)}{\kappa}\Big]^2\\
 &=\gamma F_{\rm p}\kappa\!\int\!dt\,
 \Big[F(t+\tau)-\frac{f(t+\tau)}{\kappa}\Big]^2
 \Big[\Big(1+\frac{4\Gamma_{\rm adb}}{\kappa}\Big)\langle\hat\sigma_{33}(t)\rangle
 -\frac{4gF(t)}{\kappa}\langle\hat\sigma_{13}(t)\rangle\Big].
 \end{aligned}
 \label{eq: SM closed reference}
\end{equation}
The second equality is the prodiabatic reconstruction. Thus the first line of Eq.~\eqref{eq: General G with noise} fixes this reference, while the pair term supplies the second term in Eq.~\eqref{eq: SM switch G2}. This separates the herald yield and pulse envelope from the conditional cavity memory before forming the reflectance in Sec.~\ref{sec: switch reflectance}.

\textit{Weak-drive atomic response.} To express the averages through the pulse alone, we now set $\langle\hat\sigma_{11}\rangle=1$ and drop the population term in the coherence equation. More precisely, for $f\mapsto\eta f$ we retain the coherence to order $\eta$ and the populations to order $\eta^2$; depletion and saturation enter at higher orders. Equation~\eqref{eq: SM reduced switch equations} then gives
\begin{align}
 \frac{d}{dt}\langle\hat\sigma_{13}\rangle
 &=-\Gamma_{\rm pdb}\langle\hat\sigma_{13}\rangle
 +g\Big(1+\frac{2\gamma F_{\rm p}}{\kappa}\Big)F,\nonumber\\
 \frac{d}{dt}\langle\hat\sigma_{33}\rangle
 &=-2\Gamma_{\rm pdb}\langle\hat\sigma_{33}\rangle
 +2g\Big(1+\frac{2\gamma F_{\rm p}}{\kappa}\Big)F\langle\hat\sigma_{13}\rangle.
 \label{eq: SM weak drive equations}
\end{align}
For a real pulse the coherence is real, and these equations imply
\begin{equation}
 \frac{d}{dt}\Big(\langle\hat\sigma_{33}\rangle-\langle\hat\sigma_{13}\rangle^2\Big)
 =-2\Gamma_{\rm pdb}\Big(\langle\hat\sigma_{33}\rangle-\langle\hat\sigma_{13}\rangle^2\Big).
 \label{eq: SM population identity}
\end{equation}
Both averages vanish before the pulse. Hence, within the weak-drive equations,
\begin{align}
 \langle\hat\sigma_{13}(t)\rangle
 &=g\Big(1+\frac{2\gamma F_{\rm p}}{\kappa}\Big)
 \int_0^\infty\!du\,e^{-\Gamma_{\rm pdb}u}F(t-u),\nonumber\\
 \langle\hat\sigma_{33}(t)\rangle
 &=\langle\hat\sigma_{13}(t)\rangle^2
 =g^2\Big(1+\frac{2\gamma F_{\rm p}}{\kappa}\Big)^2
 \left[\int_0^\infty\!du\,e^{-\Gamma_{\rm pdb}u}F(t-u)\right]^2.
 \label{eq: SM weak drive averages}
\end{align}
These single convolutions close Eqs.~\eqref{eq: SM switch pdb} and \eqref{eq: SM closed reference} for any real pulse. No approximation to the cavity-filtered drive $F=\chi_{\rm c}*f$ has been made here. The adiabatic forms follow by $F\to2f/\kappa$ and by dropping the factors $1+2\gamma F_{\rm p}/\kappa$.

The population equation also simplifies the Raman rate and its integral:
\begin{align}
 \kappa\langle\hat a_{\rm V}^\dagger\hat a_{\rm V}\rangle(t)
 &=\gamma F_{\rm p}\left[\langle\hat\sigma_{33}(t)\rangle
 -\frac{2}{\kappa(1+2\gamma F_{\rm p}/\kappa)}
 \frac{d}{dt}\langle\hat\sigma_{33}(t)\rangle\right],\nonumber\\
 N_{\rm V}&=\gamma F_{\rm p}\int\!dt\,\langle\hat\sigma_{33}(t)\rangle.
 \label{eq: SM raman yield}
\end{align}
The derivative integrates to zero for a pulse with vanishing initial and final excited-state population. The rate identity follows directly from the sum of the two population equations in Eq.~\eqref{eq: SM reduced switch equations}, so it also holds for the nonlinear reduced dynamics. The weak-drive approximation enters when the population is evaluated through Eq.~\eqref{eq: SM weak drive averages}.

\textit{Gaussian coincidences.} As in the End Matter we consider $f(t)=f_0e^{-(t-t_0)^2/(2\sigma^2)}$, with $f_0^2=\bar n\kappa/(\sqrt\pi\sigma)$, and we additionally take $\kappa\sigma\gg1$. To leading order in the cavity bandwidth correction, $F\simeq2f/\kappa$ and the derivative term in the Raman rate can be omitted in the pulse-weighted integrals. Squaring the convolution in Eq.~\eqref{eq: SM weak drive averages} and integrating over the herald time gives
\begin{align}
 N_{\rm V}&\simeq\sqrt\pi\,\bar n(\gamma F_{\rm p})^2
 \Big(1+\frac{2\gamma F_{\rm p}}{\kappa}\Big)^2\sigma\,\mathcal K_0(\Gamma_{\rm pdb}),\nonumber\\
 \mathcal K_0(\lambda)&=\int_0^\infty\!dp\,e^{-\lambda p}\,
 \mathrm{erf}\!\left(\frac{p}{2\sigma}\right)
 =\frac{e^{\lambda^2\sigma^2}\mathrm{erfc}(\lambda\sigma)}{\lambda},
 \label{eq: SM NV closed}\\
 \mathcal K(\lambda)&=\int_0^\infty\!dp\,e^{-\lambda p-p^2/(8\sigma^2)}\,
 \mathrm{erf}\!\left(\frac{p}{2\sigma}\right),
 \label{eq: SM gaussian kernel}\\
 P_{\rm VH}^{\rm cl}(\tau)&\simeq
 \frac{\bar n^2(\gamma F_{\rm p})^2}{\sqrt2}
 \Big(1+\frac{2\gamma F_{\rm p}}{\kappa}\Big)^2
 e^{-\tau^2/(2\sigma^2)}\,
 \mathcal K\!\left(\Gamma_{\rm pdb}+\frac{\tau}{2\sigma^2}\right),
 \label{eq: SM gaussian reference}\\
 P_{\rm VH}^{\rm pdb}(\tau)&\simeq
 \frac{\bar n^2(\gamma F_{\rm p})^2}{\sqrt2}
 \Big(1+\frac{2\gamma F_{\rm p}}{\kappa}\Big)^2
 \left[
 e^{-\tau^2/(2\sigma^2)}\mathcal K\!\left(\Gamma_{\rm pdb}+\frac{\tau}{2\sigma^2}\right)
 -\frac{2\gamma F_{\rm p}}{\kappa}e^{-\kappa\tau/2-3\tau^2/(8\sigma^2)}
 \mathcal K\!\left(\Gamma_{\rm pdb}+\frac{\tau}{4\sigma^2}\right)
 \right].
 \label{eq: SM gaussian profile}
\end{align}
Here $p$ is the sum of the two integration delays in the squared convolution. The integral over their difference gives $\mathrm{erf}(p/2\sigma)$. The Gaussian factor in $\mathcal K$ comes from weighting the Raman detection by the remaining H pulse; it is absent from the total Raman count. The two terms in Eq.~\eqref{eq: SM gaussian profile} give the closed-switch contribution and the pair-term correction to the coincidence density.

For a pulse broad also on the atomic time, $\sigma\Gamma_{\rm pdb}\gg1$, both kernels approach $1/(\sqrt\pi\sigma\lambda^2)$. Equations~\eqref{eq: SM NV closed} and \eqref{eq: SM gaussian reference} then give the expressions quoted in Eq.~\eqref{eq: switch closed form} of the End Matter,
\begin{equation}
 \boxed{\displaystyle
 N_{\rm V}\simeq\bar n\left(\frac{\gamma F_{\rm p}}{\Gamma_{\rm adb}}\right)^2,
 \qquad
 P_{\rm VH}^{\rm cl}(\tau)\simeq
 \frac{N_{\rm V}\bar n}{\sqrt{2\pi}\,\sigma}\,
 \frac{e^{-\tau^2/(2\sigma^2)}}{[1+\tau/(2\sigma^2\Gamma_{\rm pdb})]^2}.}
 \label{eq: SM gaussian profile limit}
\end{equation}
The factors $(1+2\gamma F_{\rm p}/\kappa)^2$ cancel against $\Gamma_{\rm pdb}^2$ in the yield. The reference is the herald yield times the remaining reflected flux, averaged over the herald time. Its pulse and atomic timescales are distinct from the cavity relaxation in the second term of Eq.~\eqref{eq: SM gaussian profile}. For $F_{\rm p}\gg1$, $N_{\rm V}\simeq\bar n$ in this weak-drive, broad-pulse limit.

Table~\ref{tab: SM switch values} and Fig.~\ref{fig: switch detector} use the numerical solution of Eq.~\eqref{eq: SM reduced switch equations}. The prodiabatic curve of Fig.~\ref{fig: lambda figures}(b) is instead the closed form of Eqs.~\eqref{eq: reflectance deficit} and \eqref{eq: SM gaussian reflectance}, which that solution changes by less than $10^{-3}$. At $\bar n=0.1$, saturation changes the weak-drive averages by $5\%$; at $\bar n=10^{-4}$ the two agree to $10^{-4}$. The Gaussian kernel expressions further assume a slow cavity response to the pulse. The broad-atomic-pulse limit is a separate approximation: at the figure parameters it overestimates the weak-drive zero-delay coincidence scale by about $19\%$.

Table~\ref{tab: SM switch values} compares these expressions with the full model, Eqs.~\eqref{eq: SM switch master} and \eqref{eq: SM switch Hamiltonian}, for $g/\kappa=0.1$, $\gamma/\kappa=5\times10^{-4}$, $\sigma\kappa=55$ and $\bar n=0.1$. It lists the coincidence density at zero delay, the Raman count and the rise $P_{\rm VH}(4/\kappa)/P_{\rm VH}(0)-1$. The full model has its maximum near $\tau=3.35/\kappa$. The error of prodiabatic elimination in the rise, $0.70$ percentage points, is the correction to Eq.~\eqref{eq: SM conditional displacement} discussed above; its error at zero delay is $0.8\%$. At fixed $F_{\rm p}=80$, $\sigma\gamma F_{\rm p}=2.2$ and $\bar n=0.1$, the error at zero delay is $-0.013\%$, $-0.050\%$, $-0.20\%$ and $-0.80\%$ for $\gamma F_{\rm p}/\kappa=0.005$, $0.01$, $0.02$ and $0.04$, i.e., of relative order $(\gamma F_{\rm p}/\kappa)^2$. At $\bar n=10^{-4}$, the full model gives a rise of $3.84\%$.

\begin{table}[h]
 \color{blue}
 \centering
 \begin{tabular}{lccccc}
 \hline\hline
  & $10^5P_{\rm VH}(0)/\kappa$ & $N_{\rm V}$ & rise (\%) & $R_{\rm H|V}(0)$ & $R_{\rm H|V}$ rise (\%) \\
 \hline
 Full model & 5.2793 & 0.08617 & 4.01 & 0.9266 & 6.82 \\
 adb & 5.5629 & 0.08525 & $-2.79$ & 1 & 0 \\
 First line & 5.6942 & 0.08607 & $-2.62$ & 1 & 0 \\
 pdb, Eq.~\eqref{eq: SM switch pdb} & 5.2370 & 0.08607 & 4.71 & 0.9197 & 7.53 \\
 \hline\hline
 \end{tabular}
 \caption{\rev{Coincidence density at zero delay, Raman count, conditional reflectance at zero delay, and the rises of the coincidence density and of the reflectance between $\tau=0$ and $4/\kappa$, for the pulse of Fig.~\ref{fig: lambda figures}.}}
 \label{tab: SM switch values}
\end{table}

The full model is solved in the joint space of atom and both modes and truncated at a maximum of five H photons; increasing this cutoff to six changes the profile by less than $10^{-11}$ relative. One V photon is exact, since the atom decouples after the first Raman photon. Equation~\eqref{eq: SM conditional intensity} agrees with the propagation of the conditional density matrix, and the mean counts in H, V and atomic loss, $0.011679$, $0.086165$ and $0.002155$, add up to $\bar n$. Refining the time steps and extending the pulse tails changes the normalized profiles by less than $10^{-4}$.

\rev{\textit{Existing devices.} The deficit $2\gamma F_{\rm p}/\kappa=2F_{\rm p}\gamma/\kappa$ quoted in the Letter uses $\kappa$ and $\gamma$ as energy and population decay rates, as in Eq.~\eqref{eq: SM switch master}. For the nanophotonic cavity of Ref.~\cite{Tiecke2014}, the reported cooperativity is $F_{\rm p}=(2g)^2/\kappa\gamma=7.7$ with $\kappa=2\pi\times25$~GHz and $\gamma^{-1}=26$~ns, so $\gamma/\kappa=2.4\times10^{-4}$ and $2F_{\rm p}\gamma/\kappa=0.4\%$. For the fiber-coupled microsphere of Refs.~\cite{Shomroni2014,Rosenblum2016}, the reported cooperativity $C=g^2/(\kappa_{\rm i}+\kappa_{\rm ex})\gamma=2.2$ equals $F_{\rm p}$, since doubling the amplitude rates $\kappa_{\rm i}=7.6$~MHz, $\kappa_{\rm ex}=30$~MHz and $\gamma=3$~MHz to energy and population rates leaves the ratio unchanged; then $\gamma/\kappa=0.080$ and $2F_{\rm p}\gamma/\kappa=0.35$. The expansion parameter there is $\gamma F_{\rm p}/\kappa=0.18$, so the neglected terms are of relative order $3\%$. These are chiral resonators coupled to the two propagation directions of a fiber rather than the two-mode single-sided cavity of Eq.~\eqref{eq: SM switch Hamiltonian}: the ratio transfers, the geometry does not.}

\subsection{5.4 Conditional Reflectance}\seclabel{sec: switch reflectance}{5.4}

\textit{Relation to coincidences.} The conditional reflectance measured through coincidences is
\begin{equation}
 R_{\rm H|V}(\tau)=\frac{P_{\rm VH}(\tau)}{P_{\rm VH}^{\rm cl}(\tau)},
 \label{eq: SM reflectance}
\end{equation}
where Eq.~\eqref{eq: SM closed reference} is the coincidence density expected if the H field had reached its closed-switch response at the Raman detection. This is Eq.~\eqref{eq: switch reference} of the End Matter. At a specified herald time $t$, the reflectance is the conditional H intensity divided by $\kappa[F(t+\tau)-f(t+\tau)/\kappa]^2$, as in Eq.~\eqref{eq: switch reflectance} of the Letter. Equation~\eqref{eq: SM reflectance} averages that ratio with weight
$\kappa\langle\hat a_{\rm V}^\dagger\hat a_{\rm V}\rangle(t)\,\kappa[F(t+\tau)-f(t+\tau)/\kappa]^2$.
Thus both the likelihood of the herald and the remaining closed-switch flux enter the average.

Each description uses its own Raman rate in the numerator and denominator. Adiabatic elimination and the first line of Eq.~\eqref{eq: General G with noise} give $R_{\rm H|V}=1$ identically, since their conditional intensity is already the closed-switch intensity. The pair term gives the deficit. From Eq.~\eqref{eq: SM switch G2}, its general expression at the retained prodiabatic order is
\begin{equation}
 R_{\rm H|V}(\tau)=1-
 \frac{(\gamma F_{\rm p})^2 e^{-\kappa\tau/2}}{P_{\rm VH}^{\rm cl}(\tau)}
 \int\!dt\,\langle\hat\sigma_{33}(t)\rangle F(t)
 \Big[F(t+\tau)-\frac{f(t+\tau)}{\kappa}\Big].
 \label{eq: SM filtered reflectance}
\end{equation}
Together with Eqs.~\eqref{eq: SM reduced switch equations} and \eqref{eq: SM closed reference}, this formula retains the cavity filtering of any real pulse within the validity of the elimination. The weak-drive assumption is needed only if the atomic averages are replaced by the explicit convolutions of Eq.~\eqref{eq: SM weak drive averages}. The square of the conditional field deficit and the conditional variance in Eq.~\eqref{eq: SM conditional intensity} lie beyond the order retained here.

For a finite-bandwidth pulse, the closed-switch intensity differs from the incident flux $f(t+\tau)^2/\kappa$: the cavity stores energy during the rising pulse and releases it during the falling pulse. Normalizing to the incident flux would therefore change the reference itself. For the pulse of Fig.~\ref{fig: lambda figures} it would place the first-line result at $1.027$ at zero delay and $1.037$ at $8/\kappa$, instead of unity.

\textit{Pulse-dependent factor.} Within the prodiabatic expansion, the Raman rate in the normalization of the pair term can be replaced by its leading value $\gamma F_{\rm p}\langle\hat\sigma_{33}\rangle$. The discarded denominator correction changes $R_{\rm H|V}$ only at order $(\gamma F_{\rm p}/\kappa)^2$, beyond the controlled order of prodiabatic elimination. This step retains the full filtered drive $F$ and gives
\begin{equation}
 \boxed{\begin{aligned}
 R_{\rm H|V}(\tau)&=1-\frac{2\gamma F_{\rm p}}{\kappa}
 e^{-\kappa\tau/2}\,\mathcal Q_F(\tau),\\
 \mathcal Q_F(\tau)&=
 \frac{\displaystyle\int\!dt\,\langle\hat\sigma_{33}(t)\rangle F(t)
 \left[F(t+\tau)-\frac{f(t+\tau)}{\kappa}\right]}
 {\displaystyle 2\int\!dt\,\langle\hat\sigma_{33}(t)\rangle
 \left[F(t+\tau)-\frac{f(t+\tau)}{\kappa}\right]^2}.
 \end{aligned}}
 \label{eq: SM general reflectance}
\end{equation}
The denominator must be nonzero. The population weights the herald time, $F(t)$ sets the conditional field deficit at the herald, and the bracket is the closed-switch reflected-field amplitude apart from its common factor $i\sqrt\kappa$. It includes interference with the incident field. Consequently, $\mathcal Q_F(0)$ need not equal unity when cavity filtering matters.

In the weak-drive limit, the population is given by Eq.~\eqref{eq: SM weak drive averages}; its common prefactor cancels from $\mathcal Q_F$. This makes Eq.~\eqref{eq: SM general reflectance} an explicit functional of the pulse while retaining the full cavity response.

For a pulse varying slowly over the cavity response time, $F\simeq2f/\kappa$, Eq.~\eqref{eq: SM general reflectance} reduces to
\begin{equation}
 \mathcal Q_F(\tau)\simeq\mathcal Q_f(\tau),\qquad
 \mathcal Q_f(\tau)=
 \frac{\int\!dt\,\langle\hat\sigma_{33}(t)\rangle f(t)f(t+\tau)}
 {\int\!dt\,\langle\hat\sigma_{33}(t)\rangle f(t+\tau)^2}.
 \label{eq: SM slow cavity factor}
\end{equation}
In this limit $\mathcal Q_f(0)=1$, and the weak-drive population is proportional to
$[\int_0^\infty du\,e^{-\Gamma_{\rm pdb}u}f(t-u)]^2$.

\textit{Broad pulse.} If the pulse is approximately constant over $1/\Gamma_{\rm pdb}$, then $F\simeq2f/\kappa$, the weak-drive convolution of $F$ is $2f(t)/(\kappa\Gamma_{\rm pdb})$, and
\begin{equation}
 \langle\hat\sigma_{33}(t)\rangle\simeq
 \frac{4g^2}{\kappa^2\Gamma_{\rm adb}^2}f(t)^2.
 \label{eq: SM broad population}
\end{equation}
For delays over which $f(t+\tau)\simeq f(t)$ throughout the population-weighted part of the pulse, the numerator and denominator defining $\mathcal Q_f$ in Eq.~\eqref{eq: SM slow cavity factor} coincide, giving $\mathcal Q_F(\tau)\simeq\mathcal Q_f(\tau)\simeq1$. This includes the cavity relaxation interval $\tau\sim2/\kappa$ for a smooth pulse broad on the atomic time, since $\Gamma_{\rm pdb}\ll\kappa$. We recover
\begin{equation}
 R_{\rm H|V}(\tau)\simeq1-\frac{2\gamma F_{\rm p}}{\kappa}e^{-\kappa\tau/2}.
 \label{eq: SM broad reflectance}
\end{equation}
 The broad-pulse limit is not uniform in delay: $\mathcal Q_F$ need not remain near unity when $\tau$ becomes comparable to the pulse duration.

\textit{Gaussian pulse.} For $f(t)=f_0e^{-(t-t_0)^2/(2\sigma^2)}$, each exponential convolution of the incident pulse is explicit,
\begin{equation}
 E_\lambda(t)\equiv\int_0^\infty\!du\,e^{-\lambda u}f(t-u)
 =f_0\sigma\sqrt{\frac{\pi}{2}}\,
 e^{\lambda^2\sigma^2/2-\lambda(t-t_0)}
 \mathrm{erfc}\!\left[\frac{\lambda\sigma^2-(t-t_0)}{\sqrt2\,\sigma}\right].
 \label{eq: SM emg}
\end{equation}
The filtered drive and the atomic convolution in Eq.~\eqref{eq: SM weak drive averages} are therefore
\begin{equation}
 F(t)=E_{\kappa/2}(t),\qquad
 \int_0^\infty\!du\,e^{-\Gamma_{\rm pdb}u}F(t-u)
 =\frac{E_{\Gamma_{\rm pdb}}(t)-E_{\kappa/2}(t)}{\kappa/2-\Gamma_{\rm pdb}}.
 \label{eq: SM filtered gaussian response}
\end{equation}
Using Eq.~\eqref{eq: SM weak drive averages} for the population in Eq.~\eqref{eq: SM general reflectance} then gives $\mathcal Q_F$ with only the herald-time integral left to evaluate.

In the additional limit $\kappa\sigma\gg1$, $\mathcal Q_F\simeq\mathcal Q_f$ and the Gaussian integrals already evaluated in Eq.~\eqref{eq: SM gaussian profile} give
\begin{equation}
 \mathcal Q_f(\tau)=e^{\tau^2/(8\sigma^2)}
 \frac{\mathcal K(\Gamma_{\rm pdb}+\tau/(4\sigma^2))}
 {\mathcal K(\Gamma_{\rm pdb}+\tau/(2\sigma^2))},\qquad
 R_{\rm H|V}^{\rm G}(\tau)\simeq1-\frac{2\gamma F_{\rm p}}{\kappa}
 e^{-\kappa\tau/2}\mathcal Q_f(\tau),
 \label{eq: SM gaussian reflectance}
\end{equation}
with $\mathcal K$ defined in Eq.~\eqref{eq: SM gaussian kernel}. This result applies for any $\sigma\Gamma_{\rm pdb}$ in the weak-drive, $\kappa\sigma\gg1$ regime. Since $\mathcal K$ decreases with its argument, $\mathcal Q_f(\tau)>1$ for a Gaussian at $\tau>0$: in this slow-cavity-response limit the herald-time weighting enhances the deficit relative to a bare cavity exponential.

For $\sigma\Gamma_{\rm pdb}\gg1$, the kernel limit used in Sec.~\ref{sec: switch coincidences} gives
\begin{equation}
 \mathcal Q_f(\tau)\simeq e^{\tau^2/(8\sigma^2)}
 \left[\frac{\Gamma_{\rm pdb}+\tau/(2\sigma^2)}
 {\Gamma_{\rm pdb}+\tau/(4\sigma^2)}\right]^2\simeq1,
 \qquad \tau\ll\sigma,\quad \tau\ll\sigma^2\Gamma_{\rm pdb}.
 \label{eq: SM gaussian broad factor}
\end{equation}
Multiplying Eq.~\eqref{eq: SM gaussian profile limit} by Eq.~\eqref{eq: SM broad reflectance} gives the corresponding raw coincidence density. To leading order in $\gamma F_{\rm p}/\kappa$, its initial logarithmic slope is
$\gamma F_{\rm p}-1/(\sigma^2\Gamma_{\rm pdb})$. A rise in raw coincidences therefore requires
\begin{equation}
 \sigma^2\Gamma_{\rm adb}\gamma F_{\rm p}\gtrsim1,
 \qquad\text{or}\qquad \sigma\Gamma_{\rm adb}\gtrsim1\quad(F_{\rm p}\gg1),
 \label{eq: SM rise criterion}
\end{equation}
within this broad-pulse estimate. The reflectance separates the cavity recovery from the falling pulse envelope that can obscure it in the coincidence density. At the parameters of Fig.~\ref{fig: lambda figures}, the two contributions to the estimated initial slope are $\gamma F_{\rm p}=0.040\kappa$ and $1/(\sigma^2\Gamma_{\rm pdb})=0.0076\kappa$. The prodiabatic numerical solution gives a $7.5\%$ rise of the reflectance between $\tau=0$ and $4/\kappa$, while the reference falls by $2.6\%$, leaving a $4.7\%$ rise in the coincidence density.

}

\end{document}